\RequirePackage{fix-cm}
\documentclass[twocolumn,epjc3]{svjour3}  
\smartqed  % flush right qed marks, e.g. at end of proof
\RequirePackage{graphicx}
\journalname{Eur. Phys. J. C}
\usepackage{cite}
\usepackage{amsmath}
\usepackage{amsfonts}
\usepackage{amssymb}
\usepackage{makeidx}
\usepackage{graphicx}
\usepackage{lmodern}
\usepackage{color}

\begin{document}
\title{\textbf{ Nonstatic Reissner-Nordström metric in the perturbative $f(R)$ theory: Embedding in the background of the FLRW cosmology, uniqueness of solutions, the TOV equation}}


\author{Pham Van Ky \thanksref{e,addr1,addr2}}

\thankstext{e}{e-mail: phamkyvatly AT gmail.com}

\institute{\small
\textit{Graduate university of science and technology},\\ 
\small
{\it Vietnam academy of science and technology (VAST),}\\ 
\small
{\it 18 Hoang Quoc Viet, Cau Giay, Hanoi, Viet Nam.\\}\label{addr1}
          \and
          \small
\textit{Institute of physics},\\ 
\small
\textit{Vietnam academy of science and technology (VAST)},\\ 
\small
\textit{10 Dao Tan, Ba Dinh, Hanoi, Viet Nam.\\}\label{addr2}
}

\date{Received: date / Accepted: date}
\maketitle
\begin{abstract}
This article introduces a nonstatic Reissner-Nordström metric, a metric that does not emit electromagnetic waves but can emit gravitational waves.  We first use the GR theory to study a charged spherically symmetric gravitational source (CSSGS), the obtained results are further improved in comparison with the previous studies. In particular, this article considers that the field is not necessarily static. The metric tensors $ g_{\mu\nu} $ are considered both outside and inside the gravitational source (the results show that in the first case $ g_{\mu\nu} $ are time independent, in the latter case they are time dependent). The gravitational acceleration and the event horizon of a charged black hole are investigated. The results prove that  the gravitational field is always attractive. We then use the perturbative $ f(R) $ theory to consider CSSGS. The obtained results not only correct the solution of Einstein's equation in magnitude (this will describe astronomical and cosmological quantities more accurately than Einstein's equation), but also reveal new effects. Outside the gravitational source, the metric tensors can depend on time, this makes it possible for a spherically symmetric gravitational source to emit gravitational waves (Einstein's equation cannot give this effect).  However, a spherically symmetric field still does not emit electromagnetic waves.  Next we present a new method for embedding the spherically symmetric metrics of a star (or a black hole) in the background of the FLRW cosmological. Finally, we discuss the uniqueness of the solutions of the f(R) theory. The perturbative TOV equation is also found.
\end{abstract}

\maketitle

\section{Introduction}

Einstein's equation of gravity (the GR theory) proved to be very effective when studying small-scale physical phenomena such as those in the solar system. However, when describing physical phenomena on a larger scale, such as galaxies or especially at the cosmic scale, there are many physical phenomena that GR cannot explain, such as the problems of dark matter, dark energy, inflationary epoch, etc. So we have to modify (improve) Einstein's equation to explain the physical phenomena that GR is inexplicable. The simplest extended theory of gravity (generalization of the GR theory) that is being studied with great enthusiasm today is the $f(R)$ theory.  If the Lagrangian of the gravitational field in the GR theory is $L_G = R$, it is generalized to $L_G = f(R)$ in the $f(R)$ theory. There have been many proposed models, such as the model $f(R) = R+ \lambda R^2 $ (used for the inflationary epoch), the model $ f(R)= R -\frac{ \gamma}{R} $ which could explain the problem of dark energy (the accelerated expansion of the universe today), etc. Each model can explain some of the cosmological phenomena, but none of them is perfect.\\

In an uncharged static spherically symmetric gravitational field (a static uncharged central field), we can see the exact solutions of the $ f(R) $ theory in \cite{Sebastiani:2010kv, Multamaki:2006zb, Kainulainen:2007bt, 
Shojai:2011yq, Sharif:2011uf, Erickcek:2006vf}. We can see the approximation methods for that field, but not necessarily static, in \cite{Arbuzova:2013pta, Stabile:2010zk, Capozziello:2009vr}. 
Applying some special models of $f(R)$ to study a charged central field that can be found in  \cite{Nashed:2019tuk, Nashed:2021lzq, Nashed:2021mpz, Aghamohammadi:2010yq}. These articles all work in a static field. Hitherto, there is no research that applies the  $f(R)$ theory to a non-static charged central field. This article will do that.  However, finding an exact solution for the general $f(R)$ theory (for all models, the field is not necessarily static) is very difficult.  We use the perturbative approach, only interested in the models $f(R)$ that are  slightly different from the GR theory, by applying perturbation conditions, thereby finding the consequences that it brings. Set $f(R)=R + \lambda h(R)$, where $h(R)$ is a general function that depends on the function $f(R)$, with the perturbation conditions $\mid \lambda h(R) \mid \ll \mid R \mid$. This does not mean that the curvature of space-time is small. This approach is a small correction to Einstein's theory in Lagrangian. So we named it ``the perturbative-$f(R)$ theory."     It can be applied even when the curvature of space-time is large (for example, near the event horizon of a black hole). Therefore, this approach is not the same as the approximate solution method of the field equation (which only applies when the curvature of space-time is small).  Notice that $\mid \lambda h(R) \mid \ll \mid R \mid$ does not mean  $ \lambda $  small, $ \lambda $ can still be very large, it is large or small depending on the function $ h(R)$ as long as the perturbation condition $\mid \lambda h(R) \mid \ll \mid R \mid$ is holded. This perturbation method works in the presence of gravitational source $ T_{\mu\nu} $ (unlike the other works that only work with the vacuum).   The solutions of the perturbative-$f(R)$ theory in an uncharged central field are studied in \cite{Ky:2018fer, Ky:2019gbj, VanKy:2020xxj}.  The solutions of this theory in the FLRW cosmology are studied in \cite{VanKy:2022itq}. The gravitational waves in this theory we have studied in \cite{Ky:2024lce}. In this article, we find  the solutions of the perturbative-$f(R)$ theory in a charged central field.\\

We would like to say that the GR theory is a standard theory and has been tested by numerous experiments, so the idea of a small correction to GR makes sense. This perturbative-$f(R)$ theory not only causes the gravitational wave effect of a non-static central field (Einstein's equation cannot give gravitational waves of a central field), but also  gives corrections to Einstein's theory to make the physical quantities more precise when the field is static. For instance, it will describe the precession of the planets more accurately, the precession of Mercury's orbit around the Sun or that of the star S2 orbiting the black hole Sgr A*, see details in \cite{Ky:2018fer, Ky:2019gbj, VanKy:2020xxj}. The perturbative approach also leads to some interesting effects in cosmology, see details in \cite{VanKy:2022itq}. In particular, this perturbative-$f(R)$ theory has shown that when a collapsing star approaches its event horizon, it emits very strong gravitational waves \cite{Ky:2024lce}. Thus, this theory will describe astronomical and cosmological quantities more accurately than Einstein's equation. We hope that this method can be applied to other theories, not just the $f (R)$ theory.\\

In order to apply the perturbation method to the  $f(R)$ theory, first we must  find the exact solution of the GR theory for a charged gravitational source. Using the GR equation to study a charged spherically symmetric gravitational field has been done by many authors for a long time. The first ones are Reissner (1916) and Nordström (1918), where the Reissner-Nordström metric is
\begin{eqnarray*}
ds^2=&&\left( 1-\frac{2GM}{c^2r}+\frac{{G}Q^2}{4\pi\varepsilon_0c^4r^2}\right) {dx^o}^2\nonumber\\
&&-\left( 1-\frac{2GM}{c^2r}+\frac{{G}Q^2}{4\pi\varepsilon_0c^4r^2}\right)^{-1}dr^2\nonumber\\
&&-r^2(d\theta^2+\sin^2\theta d\varphi^2).
\end{eqnarray*}  
In 1981, when considering the Reissner-Nordström metric, the authors S. M. Mahajan, A. Qadir and P. M. Valanju suggested that there is a region of space where gravity is the repulsive force. In 1983, Gr\o{}n proved the opinion of three authors, S. M. Mahajan, A. Qadir and P. M. Valanju, were wrong. That if the contribution of electromagnetic energy to the $M $ mass was taken into account, there would be  no region of space where gravity is the repulsive force \cite{Gron}. A. Qadir immediately argued that Gr\o{}n's proof was incorrect, that there still exists a region of space where the gravitational field generates a force of  repulsion \cite{Asghar}. We will rework these problems in detail. If the gravitational field generates a force of attraction with a positive gravitational acceleration, then the gravitational field generates a force of repulsion with a negative gravitational acceleration. Both of the articles \cite{Gron} and \cite{Asghar} misdefined the mathematical expression of the gravitational acceleration. These two articles also did not provide the  details of the mathematical expression of mass $M$ in the Reissner-Nordström metric. In 1982, another author studied the metric of a charged gravitational source but in the form of a point, \cite{clp}. There are some interesting properties that do not appear because of the point-like source of gravity, such as the absence of black hole or the dependence of the metric tensor  $g_{\mu\nu}$ on the radius of the gravitational source and the charge distribution inside the gravitational source has been neglected. Recently, some authors have also studied the spherically symmetric field  of a charged gravitational source such as \cite{Rashida, Marsh:2007ib}. However, \cite{Rashida} did not consider the meaning of mass $M$ in the Reissner-Nordström metric, nor did it consider the event horizon of a charged black hole. In \cite{Marsh:2007ib} the meaning of mass $ M $ was considered, but when considering the gravitational acceleration of a neutral particle in the Reissner-Nordström metric, the mathematical expression of the acceleration due to gravity was also not defined correctly, which in turn led  to the conclusion that there exists a region of space where the gravitational field is repulsive. All the mentioned articles only consider the metric that is outside of the gravitational source and in a static field.\\

With all that said, this article will re-work the problem of finding the metric of a charged spherically symmetric gravitational source, but the field is not necessarily static. Furthermore, the metric tensors inside the charged gravitational source are also considered (this might make some sense in physics if we consider the physical processes that take place inside a star or a black hole). We obtained the results as shown in the abstract. The uniqueness of the solutions of the $ f(R) $ theory  is discussed in the 6 section. This is probably a new topic for us to study further. The TOV equation found in this perturbative f(R) theory can be used even when the gravitational source is charged, it is explicit and simple.\\

G. C. McVittie (1933) was probably the first to study the embedding of the Schwarzschild metric in the background of the  universe, \cite{McVittie}. In 2004, C.~J.~Gao and S.~N.~Zhang studied the static Reissner-Nordström metric embedding in the background of the FLRW  universe, \cite{Gao:2004cr}. To achieve embedding results, both of these articles impose the condition that the metric after embedding must satisfy $ G_{01}=0 $ ($ G_{\mu\nu} $ is the Einstein tensor). Many other authors have also offered different embedding methods by imposing certain conditions, thereby getting different embedding results, refer to the article \cite{Arik:2014qua}. In addition, the analysis of the meanings and effects of the metric after embedding can be found in \cite{Kaloper:2010ec, Lake:2011ni, Nandra:2011ug, daSilva:2012nh}. In this article, we found that there is a new approach, the result of the embedding  appears very naturally without any conditions being imposed. The calculations in this approach are much simpler than in previous methods.\\

In this article the following conventions are used:
\begin{itemize}
\item Metric signature in Minkowski space: ($ +,  -,  -,  - $), that is, 
the infinitesimal distance is calculated as 
\begin{align*}
ds^2&=\eta_{\mu\nu}dx^{\mu}dx^{\nu}=c^2dt^2-dx^2-dy^2-dz^2.
\end{align*} 
\item  Riemann curvature tensor: 
\begin{eqnarray*}
R^{\alpha}_{~\mu\beta\nu}=\frac{\partial \Gamma^\alpha_{\mu\beta}}{\partial x^\nu} - 
\frac{\partial \Gamma^\alpha_{\mu\nu}}{\partial x^\beta} + 
\Gamma^\alpha_{\sigma\nu}\Gamma^\sigma_{\mu\beta} -  
\Gamma^\alpha_{\sigma\beta}\Gamma^\sigma_{\mu\nu}.
\end{eqnarray*}
\item Rank-2 curvature tensor (Ricci tensor): $R_{\mu\nu}=R^\alpha_{~\mu\alpha\nu}.$
\item Scalar curvature:  $ R=g^{\mu\nu}R_{\mu\nu} $.
\item Energy-momentum tensor of a macroscopic object: 
\begin{eqnarray*}
T_{\mu\nu}
=\frac{1}{c^2}(\varepsilon + p)u_\mu u_\nu - pg_{\mu\nu}, 
\end{eqnarray*}
where $ u^\mu 
= \frac{dx^\mu}{d\tau}=c\frac{dx^\mu}{ds} $, while $ \varepsilon$ and $p$ 
are the energy density and the pressure, respectively.
\item The Einstein equation      
\begin{align*} 
R_{\mu\nu}-\frac{1}{2}g_{\mu\nu}R=-\frac{8\pi{G}}{c^4}T_{\mu\nu}.
\end{align*}
\end{itemize}	

\section{The Maxwell equations in a spherically symmetric gravitational field}
In order  to solve the equations of gravity theories (not just the GR theory) in the spherically symmetric field of a charged gravitational source, we must know the energy-momentum tensor of the electromagnetic field in the gravitational field. As we already know, the action of matter-radiation in curved space-time is (\cite{L2})
\begin{equation}
S^{(me)}=\frac{1}{c}\int L^{(me)} \sqrt{-g}d^4x. \label{S1}
\end{equation}
Where $ L^{(me)} $ is Lagrangian of matter-radiation. The energy-momentum tensor of matter-radiation in curved spacetime can be derived by varying the action \eqref{S1} with respect to the metric tensor $ g_{\mu\nu} $ (\cite{L2})
\begin{align}
T_{\mu\nu}=&\frac{2}{\sqrt{-g}}\left[ \frac{\partial{(L^{(me)}\sqrt{-g})}}{\partial{g^{\mu\nu}}}-\frac{\partial}{\partial{x^{\alpha}}}\left( \frac{\partial{(L^{(me)}\sqrt{-g})}}{\partial{\frac{\partial{g^{\mu\nu}}}{\partial{x^{\alpha}}}}}\right) \right]\nonumber\\
=&-L^{(me)}g_{\mu\nu}+2\frac{\partial{L^{(me)}}}{\partial{g^{\mu\nu}}}\nonumber\\
&-\frac{2}{\sqrt{-g}}\frac{\partial}{\partial{x^{\alpha}}}\left( \frac{\partial{(L^{(me)}\sqrt{-g})}}{\partial{\frac{\partial{g^{\mu\nu}}}{\partial{x^{\alpha}}}}}\right) . \label{T}
\end{align}
From \eqref{T}, if $L^{(me)}=L^{(e)} $ with the Lagrangian of the electromagnetic field $ L^{(e)} $ is
\begin{equation}
L^{(e)}=-\frac{1}{4}\varepsilon_0 c^2F^{\alpha\beta}F_{\alpha\beta}=-\frac{1}{4}\varepsilon_0 c^2g^{\alpha\sigma}g^{\beta\rho}F_{\sigma\rho}F_{\alpha\beta}, \label{T1}
\end{equation}
we obtain the energy-momentum tensor of the electromagnetic field in the gravitational field, which is (\cite{L2})
\begin{equation}
T^{(e)}_{\mu\nu}=\varepsilon_0c^2\left(-F_{\mu\alpha}F_{\nu\beta}g^{\alpha\beta}+\frac{1}{4}F_{\alpha\beta}F^{\alpha\beta}g_{\mu\nu} \right). \label{TE}
\end{equation}
Where $ \frac{1}{4\pi\varepsilon_0}=9\times10^9~Nm^2/C^2 $, 
\begin{equation}
F_{\mu\nu}=\nabla_\mu A_\nu-\nabla_\nu A_\mu=\partial_\mu A_\nu-\partial_\nu A_\mu, \label{F}
\end{equation}
 $ \nabla_\mu $ is the covariant derivative. From \eqref{TE}, it is easy to see that
\begin{equation}
T^{(e)}=g^{\mu\nu}T^{(e)}_{\mu\nu}=T^{\mu(e)}_{~~\mu}=0. \label{TEa}
\end{equation}
As we already know, for every vector $ A_\mu $, we have
\begin{eqnarray}
&&\nabla_\alpha\nabla_\beta A_\lambda-\nabla_\beta\nabla_\alpha A_\lambda=-A_\sigma R^\sigma_{~\lambda\beta\alpha}, \label{F3}\\
&&R^\sigma_{~\lambda\beta\alpha}+R^\sigma_{~\alpha\lambda\beta}+R^\sigma_{~\beta\alpha\lambda}=0, \label{F4}
\end{eqnarray}
so from \eqref{F} it is easy to see that
\begin{eqnarray}
\nabla_\alpha F_{\mu\nu}+\nabla_\nu F_{\alpha\mu}+\nabla_\mu F_{\nu\alpha}=0,\label{F1}
\end{eqnarray}
it follows that
\begin{equation}
\partial_\alpha F_{\mu\nu}+\partial_\nu F_{\alpha\mu}+\partial_\mu F_{\nu\alpha}=0. \label{F2}
\end{equation}

The equation \eqref{F1} (also the equation \eqref{F2}) is the first pair of Maxwell's equations. In order  to find the second pair of Maxwell's equations we do the following: The proper time interval $ d\tau $ (the real time) is (\cite{L2})
\begin{equation}
d\tau = \sqrt{g_{00}}dt, \label{tau1}
\end{equation}
 $ d\tau $ is the time measured by an observer located at the given point. The equation \eqref{tau1} can be derived by the equation $ ds=cd\tau $ with $ dx=dy=dz=0 $. The spatial distance element $ dl $ (the proper distance or the real distance) (\cite{L2})
 \begin{equation}
 dl^2=\gamma_{ij}dx^idx^j, \label{dl}
 \end{equation}
 with $ i,~ j= (1;~ 2;~ 3) $ and
 \begin{equation}
 \gamma_{ij}=-g_{ij}+\frac{g_{0i}g_{0j}}{g_{00}} \label{gamma}
 \end{equation}
 is the three-dimensional metric tensor. In a spherically symmetric field, from \eqref{gamma} it is easy to see that $ \gamma_{ij}=-g_{ij} $ (because $ g_{0i}=0 $), it means that
 \begin{equation}
 ds^2 = g_{00}{dx^0}^2 - dl^2=c^2d\tau^2 - dl^2. \label{ds}
 \end{equation}
From \eqref{dl}, it is easy to see that the spatial volume element (the proper volume or the real volume) is 
 \begin{equation}
 dV=\sqrt{\gamma}dx^1dx^2dx^3, \label{dV}
 \end{equation}
 with $\gamma=\gamma_{11}\gamma_{22}\gamma_{33}$. From \eqref{gamma}, it follows that 
\begin{equation}
g=-g_{00}\gamma. \label{F5}
\end{equation}
The charge is (\cite{L2})
\begin{equation}
de=\rho dV=\rho \sqrt{\gamma}dx^1dx^2dx^3, \label{F5a}
\end{equation}
 so the charge density is 
\begin{eqnarray}
\rho=&&\frac{1}{\sqrt{\gamma}}\sum_{a} e_a \delta(\overrightarrow{x}-\overrightarrow{x_a})\nonumber\\
=&&\frac{1}{\sqrt{\gamma}}\sum_a e_a\delta(x^1-x^1_a)\delta(x^2-x^2_a)\delta(x^3-x^3_a). \label{F6}
\end{eqnarray}
Notice that
\begin{eqnarray}
\delta(\overrightarrow{r})&&=\delta(x)\delta(y)\delta(z) \nonumber\\
&&=\delta(r\sin\theta\cos\varphi)\delta(r\sin\theta\sin\varphi)\delta(r\cos\theta) \nonumber\\
&&=\frac{1}{r^2\sin\theta}\delta(r)\delta(\theta)\delta(\varphi). \label{F6a}
\end{eqnarray}
It is easy to see the correctness of the formula \eqref{F6a} if using the transformation $ dxdydz=r^2\sin\theta drd\theta d\varphi $, and from that we see 
\begin{eqnarray*}
&&\int \delta(x )\delta(y)\delta(y)dxdydz\nonumber\\
&&=\int \frac{\delta(r)\delta(\theta)\delta(\varphi)r^2\ sin\theta drd\theta d\varphi}{r^2\sin\theta} =1
\end{eqnarray*} 
is satisfied.\\[2mm]
We have 
\begin{eqnarray*}
de dx^\mu=&& \rho \sqrt{\gamma}dx^1dx^2dx^3dt\frac{dx^\mu}{dt}\nonumber\\
=&&\rho \frac{\sqrt{\gamma}}{\sqrt{-g}}\sqrt{-g}dx^1dx^2dx^3dt\frac{dx^\mu}{dt}.
\end{eqnarray*}
On the other hand, $ de dx^\mu $ is a vector and $ \sqrt{-g}dx^1dx^2dx^3dt $ is the invariant element , so  $ \rho \frac{\sqrt{\gamma}}{\sqrt{-g}}\frac{dx^\mu}{dt}=\rho\frac{dx^\mu}{\sqrt{g_{00}}dt} $ is a vector. It follows that the current four-vector is defined by the expression (\cite{L2})
\begin{equation}
j^\mu=\rho\frac{dx^\mu}{\sqrt{g_{00}}dt}. \label{F7}
\end{equation}
In the special theory of relativity, we already know the second pair of Maxwell's equations is $ \partial_\nu F^{\mu\nu}=-\frac{1}{\varepsilon_0c^2}j^\mu $. In the curved space-time, $\partial_\mu$ is replaced by $ \nabla_\mu $, so 
\begin{equation}
\nabla_\nu F^{\mu\nu}=-\frac{1}{\varepsilon_0c^2}j^\mu. \label{F8}
\end{equation}
The equation \eqref{F8} can be expressed as
\begin{equation}
\frac{1}{\sqrt{-g}}\partial_\nu(\sqrt{-g}F^{\mu\nu})=-\frac{1}{\varepsilon_0c^2}j^\mu. \label{F9}
\end{equation}\\

\textit{Spherically symmetric spacetime:} Now we consider the equations in a spherically symmetric field. For simplicity, the spherical coordinates ($r, \theta, \varphi $) will be used. In a spherically symmetric field, as we already know, both inside and outside the spherically symmetric charged gravitational source, the magnetic induction $B$ is zero.  So the vector potentials $ A_1, A_2, A_3 $ are all zero, only the scalar potential $ A_0$ remains. Outside the gravitational source, apply Gauss's law we will see that $ A_0 $ depends only on the variable $ r $, $ A_0=A_0(r) $ (we can also prove this by solving Einstein's equation (see below)). Inside the gravitational source $ A_0=A_0 (r, t) $. Thus, there is only one non-zero component of the tensor $ F_{\mu\nu} $ which is
\begin{eqnarray}
F_{01}=-F_{10}=-\frac{\partial A_0(r, t)}{\partial r}. \label{F10}
\end{eqnarray}
Since the spherically symmetric field $ g_{\mu\nu}=0 $ when $ \mu\neq \nu $, there is also only one non-zero component of the tensor $ F^{\mu\nu} $ which is
\begin{eqnarray}
F^{01}=-F^{10}=-g^{00}g^{11}\frac{\partial A_0(r, t)}{\partial r}. \label{F11}
\end{eqnarray}
From here, we obtain the non-zero components of the  \eqref{TE} tensor as
\begin{eqnarray}
&&T^{0(e)}_{~0}=g^{00}T^{(e)}_{00}\nonumber\\
&&=\varepsilon_0 c^2g^{00}\left(-F_{01}F_{01}g^{11}+\frac{1}{4}F_{01}F^{01}g_{00}+\frac{1}{4}F_{10}F^{10}g_{00} \right) \nonumber\\
&&= -\frac{\varepsilon_0 c^2}{2}\left( \frac{\partial A_0(r, t)}{\partial r}\right)^2g^{00}g^{11}, \label{TE1}
\end{eqnarray}
\begin{eqnarray}
&&T^{1(e)}_{~1}=g^{11}T^{(e)}_{11}\nonumber\\
&&=\varepsilon_0 c^2g^{11}\left(-F_{10}F_{10}g^{00}+\frac{1}{4}F_{01}F^{01}g_{11}+\frac{1}{4}F_{10}F^{10}g_{11} \right) \nonumber\\
&&= -\frac{\varepsilon_0 c^2}{2}\left( \frac{\partial A_0(r, t)}{\partial r}\right)^2g^{00}g^{11}, \label{TE2}
\end{eqnarray}
\begin{eqnarray}
T^{2(e)}_{~2}=g^{22}T^{(e)}_{22}=&&\varepsilon_0 c^2g^{22}\left(\frac{1}{4}F_{01}F^{01}g_{22}+\frac{1}{4}F_{10}F^{10}g_{22} \right) \nonumber\\
=&&\frac{\varepsilon_0 c^2}{2}\left( \frac{\partial A_0(r, t)}{\partial r}\right)^2g^{22}g^{00}g^{11}g_{22}\nonumber\\
=&&\frac{\varepsilon_0 c^2}{2}\left( \frac{\partial A_0(r, t)}{\partial r}\right)^2g^{00}g^{11},\label{TE3}
\end{eqnarray}
\begin{eqnarray}
T^{3(e)}_{~3}=g^{33}T^{(e)}_{33}=&&\varepsilon_0 c^2g^{33}\left(\frac{1}{4}F_{01}F^{01}g_{33}+\frac{1}{4}F_{10}F^{10}g_{33} \right) \nonumber\\
=&&\frac{\varepsilon_0 c^2}{2}\left( \frac{\partial A_0(r, t)}{\partial r}\right)^2g^{33}g^{00}g^{11}g_{33}\nonumber\\
=&&\frac{\varepsilon_0 c^2}{2}\left( \frac{\partial A_0(r, t)}{\partial r}\right)^2g^{00}g^{11}. \label{TE4}
\end{eqnarray}
The first pair of Maxwell's equations were automatically satisfied in all cases. The second pair of Maxwell's equations \eqref{F9} are
\begin{equation}
\frac{1}{\sqrt{-g}}\frac{\partial}{\partial r}\left( \sqrt{-g}g^{00}g^{11}\frac{\partial A_0(r, t)}{\partial r}\right) =\frac{\rho}{\varepsilon_0 c \sqrt{g_{00}}}, \label{MW1}
\end{equation}
\begin{equation}
\frac{1}{\sqrt{-g}}\frac{\partial}{\partial t}\left( \sqrt{-g}g^{00}g^{11}\frac{\partial A_0(r, t)}{\partial r}\right) =-\frac{\rho v_r}{\varepsilon_0 c \sqrt{g_{00}}}. \label{MW2}
\end{equation}
Where $ v_r = \frac{dr}{dt} $ is the velocity of $ \rho $ in the direction $ r $.

\section{Solutions of the GR theory outside of a charged spherically symmetric gravitational source}
We now consider the matter-radiation system and the gravitational field. If the Lagrangian of the gravitational field is ${L}_G = R$ and that of matter-radiation is ${L}^{(me)}$, then the action of the system will be
\begin{align}
S&=S_G + S^{(me)}\nonumber\\
&=\frac{c^3}{16\pi{G}}\int{R{\sqrt{-g}} {d^4x}}+\frac{1}{c}\int{{L}^{(me)}{\sqrt{-g}}{d^4x}}. 
\label{1}
\end{align}
The Einstein equation  be obtained in this case  (\cite{L2, Weinberg})    
\begin{align} 
R_{\mu\nu}-\frac{1}{2}g_{\mu\nu}R=-\frac{8\pi{G}}{c^4}T_{\mu\nu}.\label{2}
\end{align}
Where $T_{\mu\nu}$ is the energy-momentum tensor of matter-radiation in \eqref{T}. Taking traces of the equation \eqref{2}, we get  
\begin{align}
R=\frac{8\pi{G}}{c^4}T \label{R}
\end{align}
with $T=T^\mu_{~\mu}$, and the equation \eqref{2} becomes
\begin{align} 
R_{\mu\nu}=-\frac{8\pi{G}}{c^4}\left (T_{\mu\nu}-{1\over 2} g_{\mu\nu}T\right). \label{RT}
\end{align} 
Notice that
\begin{eqnarray}
T^\mu_{~\nu}=T^{\mu(m)}_{~\nu}+T^{\mu(e)}_{~\nu}, \label{me1}
\end{eqnarray}
where $T^{\mu(m)}_{~\nu} $ is the energy-momentum tensor of matter, $ T^{\mu(e)}_{~\nu} $ is the energy-momentum tensor  of electromagnetic field. We use the spherically symmetric metric in the shape of the Schwarzschild metric (\cite{L2}), 
\begin{align}
ds^2=e^{u(r,t)}{dx^o}^2-e^{v(r,t)}dr^2-r^2(d\theta^2+\sin^2\theta d\varphi^2), \label{11}
\end{align}
with the following non-zero metric elements
\begin{align*}
g_{00}=e^{u(r,t)}, ~g_{11}= -e^{v(r,t)},\\
~g_{22}= -r^2 ,~ g_{33}=-r^2\sin^2\theta.
\end{align*}
Using metric \eqref{11}, we have  (\cite{L2}) 
\begin{align}
R^0_{~0}-\frac{1}{2}R=-e^{-v(r,t)}\left[ \frac{v'(r,t)}{r}-\frac{1}{r^2}\right] -\frac{1}{r^2},  \label{28}\\
R^1_{~1}-\frac{1}{2}R=e^{-v(r,t)}\left[ \frac{u'(r,t)}{r}+\frac{1}{r^2}\right] -\frac{1}{r^2}. \label{29}
\end{align}
where $ v'(r,t)=\frac{\partial v(r,  t)}{\partial r} $. The  \eqref{2} equations only two independent equations are (\cite{L2}) 
\begin{align}
-e^{-v(r,t)}\left[ \frac{v'(r,t)}{r}-\frac{1}{r^2}\right] -\frac{1}{r^2}=-\frac{8\pi{G}}{c^4}T^0_{~0},  \label{28a}\\
e^{-v(r,t)}\left[ \frac{u'(r,t)}{r}+\frac{1}{r^2}\right] -\frac{1}{r^2}=-\frac{8\pi{G}}{c^4}T^1_{~1}.  \label{29a}
\end{align}
Set
\begin{equation}
v(r, t)= -\ln\left[1+\frac{C(r, t)}{r} \right], \label{v1}
\end{equation}
the equation \eqref{28a} becomes
\begin{eqnarray}
\frac{C'(r, t)}{r^2}=-\frac{8\pi{G}}{c^4}T^0_{~0}. \label{C1}
\end{eqnarray}
It follows that (notice when $ r\longrightarrow 0 $ then $ C(r)\longrightarrow 0 $, otherwise the left hand side of \eqref{28a} will go to infinity)
\begin{eqnarray}
&&C(r, t)=-\frac{8\pi{G}}{c^4}\int_0^r T^0_{~0} r^2dr, \label{C2a}\\
&&=-\frac{8\pi{G}}{c^4}\int_0^r T^{0(m)}_{~0} r^2dr-\frac{8\pi{G}}{c^4}\int_0^r T^{0(e)}_{~0} r^2dr. \label{C2}
\end{eqnarray}
Outside of the charged gravitational source, $T^{\mu(m)}_{~\nu}=0 $ then $T^\mu_{~\nu}=T^{\mu(e)}_{~\nu} $. On the other hand, from \eqref{TE1} and \eqref{TE2} we see that $T^{0(e)}_{~0}=T^{1(e)}_{~1} $. So outside of the charged gravitational source we have $T^{0}_{~0}=T^{1}_{~1} $, combining this with \eqref{28a} and \eqref{29a} we obtain
\begin{equation}
u(r, t) = -v(r, t). \label{me2}
\end{equation}
It follows that
\begin{equation}
g_{00}g_{11}=g^{00}g^{11}=-1. \label{me3}
\end{equation}
On the other hand (\cite{L2})
\begin{equation}
R^1_{~0}=\frac{e^{-v(r, t)}}{r}\frac{\partial v(r, t)}{\partial ct},
\end{equation}
from \eqref{2} we have $ R^1_{~0}= -\frac{8\pi{G}}{c^4}T^1_{~0}$, outside of the gravitational source $ T^{\mu (m)}_{~\nu}=0 $ and $ T^{1(e)}_{~0}=0 $. Therefore,
\begin{equation}
\frac{\partial v(r, t)}{\partial t}=0. \label{me4}
\end{equation}
Thus, the  metric tensors of the spherically symmetric field outside a charged gravitational source will be  time-independent  (stationary state)  with $ u(r, t)=u(r) $, $ v(r, t)=v(r) $ and $ u(r)=-v(r) $. Combining \eqref{me3} with \eqref{11} we can rewrite \eqref{MW1} and \eqref{MW2} as (notice that outside the gravitational source the charge density $ \rho=0 $)
\begin{eqnarray}
-\frac{1}{r^2}\frac{\partial}{\partial r}\left( r^2\frac{\partial A_0(r, t)}{\partial r}\right) =0, \label{me6}\\
-\frac{1}{r^2}\frac{\partial}{\partial t}\left( r^2\frac{\partial A_0(r, t)}{\partial r}\right) =0. \label{me7}
\end{eqnarray}
From \eqref{me6} and \eqref{me7} we have (select the condition when $ r\longrightarrow \infty $ then $ A_0(\infty, t)\longrightarrow 0 $)
\begin{equation}
A_0(r, t)= A_0(r)=-\frac{C_1}{r}. \label{me8}
\end{equation}
Where $C_1$ is a constant. Thus, outside the gravitational source, we see that, the potential functions of the electromagnetic field  in a spherically symmetric field  are also time-independent  (stationary state). We choose $C_1=-\frac{Q}{4\pi \varepsilon_0 c} $, where $Q $ is the total charge of the gravitational source, then
\begin{equation}
A_0(r)=\frac{Q}{4\pi \varepsilon_0 cr}. \label{me9}
\end{equation}
Set
\begin{equation}
E=-c\frac{\partial A_0(r)}{\partial r} = \frac{Q}{4\pi \varepsilon_0 r^2}, \label{me10}
\end{equation}
$E$ is called the electric field strength. Outside the gravitational source, using \eqref{me10},   \eqref{me9}, and \eqref{me3} we rewrite the formulas \eqref{TE1}--\eqref{TE4} as
\begin{eqnarray}
&&T^{0(e)}_{~0}=T^{1(e)}_{~1}=-T^{2(e)}_{~2}=-T^{3(e)}_{~3}  \label{me13}\\
&&=\frac{\varepsilon_0 c^2}{2}\left( \frac{\partial A_0(r)}{\partial r}\right)^2= \frac{1}{2}\varepsilon_0\left(\frac{Q}{4\pi \varepsilon_0 r^2} \right)^2=\frac{1}{2}\varepsilon_0 E^2, \nonumber
\end{eqnarray}
with $ T^{0(e)}_{~0}= \frac{1}{2}\varepsilon_0 E^2$ is the energy density of the electric field. It is easy to see that, inside a spherically symmetric gravitational source, we can take the energy density of the electric field at any position $ r $ as $T^{0(e)}_{~0}(r, t)$ with
\begin{equation}
T^{0(e)}_{~0}(r, t)=\frac{1}{2}\varepsilon_0\left(\frac{Q(r, t)}{4\pi \varepsilon_0 r^2} \right)^2=\frac{1}{2}\varepsilon_0 E^2(r, t), \label{meg1}
\end{equation}
where $E(r, t) $ is the electric field strength at position $ r $ and $ Q(r, t) $ is the total charge presented inside the sphere of radius $ r $.\\

When considering outside the gravitational source with $ r > R_0 $ ($ R_0=R_0(t) $ is the radius of the gravitational source), then $ T^{0(m)}_{~0}=0 $. Thus, in the formula \eqref{C2}, the first integral can be taken from $0$ to $R_0$. Set
\begin{equation}
\int_0^{R_0}T^{0(m)}_{~0} d\frac{4\pi r^3}{3}=Mc^2, \label{C4}
\end{equation}
then \eqref{C2} becomes (using \eqref{meg1})
\begin{eqnarray}
C(r)=&&-\frac{2GM}{c^2}-\frac{8\pi{G}}{c^4}\int_0^r \frac{1}{2}\varepsilon_0E^2(r, t) r^2dr, \label{C3a}\\
=&&-\frac{2GM}{c^2}-\frac{8\pi{G}}{c^4}\int_{R_0}^r \frac{1}{2}\varepsilon_0E^2(r) r^2dr\nonumber\\
&&-\frac{8\pi{G}}{c^4}\int_0^{R_0} \frac{1}{2}\varepsilon_0E^2(r, t) r^2dr, \label{C3}\\
=&&-\frac{2GM}{c^2}+\frac{{G}Q^2}{4\pi\varepsilon_0c^4}\left( \frac{1}{r}-\frac{1}{R_0}\right)\nonumber\\
&& -\frac{8\pi{G}}{c^4}\int_0^{R_0} \frac{1}{2}\varepsilon_0E^2(r, t) r^2dr. \label{C5}
\end{eqnarray}
Now let's look at some cases:\\
\textit{+ The uniformly distributed charge inside the gravitational source:}  In this case, according to \eqref{meg1} we can write
\begin{equation}
E(r, t)=\frac{Q(r, t)}{4\pi\varepsilon_0r^2}=\frac{\frac{Qr^3}{R_0^3}}{4\pi\varepsilon_0r^2}=\frac{Qr}{4\pi\varepsilon_0R_0^3}, \label{dtd1}
\end{equation}
then \eqref{C5} becomes
\begin{equation}
C(r)=-\frac{2GM}{c^2}+\frac{{G}Q^2}{4\pi\varepsilon_0c^4}\left( \frac{1}{r}-\frac{6}{5R_0}\right). \label{dtd2}
\end{equation}
\textit{+ The surface charge:} When the charge is concentrated on the surface of the gravitational source then $ E=0 $ for $ 0 < r < R_0 $. The formula \eqref{C5} becomes
\begin{equation}
C(r)=-\frac{2GM}{c^2}+\frac{{G}Q^2}{4\pi\varepsilon_0c^4}\left( \frac{1}{r}-\frac{1}{R_0}\right). \label{dtbm1}
\end{equation}
\textit{+ The uniformly distributed charge in the interval $ r $ with $ R_e < r < R_0 $:} In this case $ E=0 $ if $ r < R_e $ and the formula \eqref{C5} becomes
\begin{eqnarray}
C(r)=-\frac{2GM}{c^2}+\frac{{G}Q^2}{4\pi\varepsilon_0c^4}\left( \frac{1}{r}-\frac{1}{R_0}\right)\nonumber\\ -\frac{8\pi{G}}{c^4}\int_{R_e}^{R_0} \frac{1}{2}\varepsilon_0E^2(r, t) r^2dr, \label{vk1}
\end{eqnarray}
when $ R_e < r < R_0 $ then 
\begin{equation}
E(r, t)=\frac{Q(r, t)}{4\pi\varepsilon_0r^2}=\frac{\frac{Q\left( r^3-R_e^3\right)}{R_0^3-R_e^3}}{4\pi\varepsilon_0r^2}=\frac{Q\left( r-\frac{R_e^3}{r^2}\right) }{4\pi\varepsilon_0\left( R_0^3-R_e^3\right)}, \label{vk2}
\end{equation}
\eqref{vk1} becomes
\begin{eqnarray}
&&C(r)=\nonumber\\
&&-\frac{{G}Q^2\left[ \frac{1}{5}\left(R_0^5-R_e^5 \right)-R_e^3\left(R_0^2-R_e^2 \right)-R_e^6\left( \frac{1}{R_0}-\frac{1}{R_e}\right) \right]}{4\pi\varepsilon_0c^4\left(R_0^3-R_e^3 \right)^2 }\nonumber\\
&&-\frac{2GM}{c^2}+\frac{{G}Q^2}{4\pi\varepsilon_0c^4}\left( \frac{1}{r}-\frac{1}{R_0}\right). \label{vk3}
\end{eqnarray}
\\

Thus, from \eqref{C5}, \eqref{me2} and \eqref{11} we have the metric of a spherically symmetric field outside the charged gravitational source is 
\begin{align}
ds^2=&\left( 1-\frac{2GM_f}{c^2r}+\frac{{G}Q^2}{4\pi\varepsilon_0c^4r^2}\right) {dx^o}^2\nonumber\\
&-\left( 1-\frac{2GM_f}{c^2r}+\frac{{G}Q^2}{4\pi\varepsilon_0c^4r^2}\right)^{-1}dr^2\nonumber\\
&-r^2(d\theta^2+\sin^2\theta d\varphi^2). \label{k1}
\end{align}
Where (see the formula \eqref{C5})
\begin{equation}
M_f = M + M_{(e)outs}+ M_{(e)ins}, \label{k2}
\end{equation}
with $ R_0=R_0(t) $ is the radius of the gravitational source and
\begin{eqnarray}
&&M_{(e)outs}=\frac{Q^2}{8\pi\varepsilon_0c^2R_0}, \label{outs}\\
&&M_{(e)ins}=\frac{4\pi}{c^2}\int_0^{R_0} \frac{1}{2}\varepsilon_0E^2(r, t) r^2dr. \label{k2fr}
\end{eqnarray}
We call $M_f$ as the total mass (the sum of the masses of the matter, the gravitational field and the electromagnetic field). Where $M_{(e)outs}$ is the mass of the electric field outside the gravitational source and $M_{(e)ins}$ is the mass of the electric field inside the gravitational source. Indeed, from \eqref{meg1} we easily see
\begin{align}
M_f&=M + \frac{Q^2}{8\pi\varepsilon_0c^2R_0}+ \frac{4\pi}{c^2}\int_0^{R_0} \frac{1}{2}\varepsilon_0E^2(r, t) r^2dr\nonumber\\
&=M+ \frac{1}{c^2}\int^\infty_{R_0} T^{0(e)}_{~0} d\frac{4\pi r^3}{3}+\frac{1}{c^2}\int^{R_0}_0 T^{0(e)}_{~0} d\frac{4\pi r^3}{3}\nonumber\\
%&=M+ \frac{1}{c^2}\int^\infty_0 T^{0(e)}_{~0} d\frac{4\pi r^3}{3}\nonumber\\
&=M+M_{(e)} \label{Mf}\\
&=\mbox{constant} \nonumber, 
\end{align}
($M_f$=constant, which will be proved below). Where
\begin{equation}
M_{(e)}=\frac{1}{c^2}\int^\infty_0 T^{0(e)}_{~0} d\frac{4\pi r^3}{3}. \label{Me}
\end{equation}
Thus, the mass of matter $M$ is in \eqref{C4}, and the mass of the electric field $M_{(e)} $ is in \eqref{Me}. As mentioned below the formula \eqref{me4}, the  metric tensors outside the gravitational source are time-independent. Therefor, from \eqref{k1} we see $M_f=\mbox{constant}$. When the radius of the gravitational source changes with time $ R_0=R_0(t) $, then $ M_{(e)} $ and $ M $ will change, but, their sum $ M_f $ will not change. This shows the conservation of total energy.  \\[2mm]
\textit{+ If the charge is uniformly distributed inside the gravitational source,} then according to \eqref{dtd2}
\begin{equation}
M_{(e)ins}=\frac{Q^2}{40\pi\varepsilon_0c^2R_0}, \label{k4fr}
\end{equation}
\begin{equation}
M_f = M + \frac{3Q^2}{20\pi\varepsilon_0c^2R_0}. \label{k4}
\end{equation}
\textit{+ If the charge is concentrated on the surface of the gravitational source,} then according to \eqref{dtbm1}
\begin{equation}
M_{(e)ins}=0, \label{k3fr}
\end{equation}
\begin{equation}
M_f = M + \frac{Q^2}{8\pi\varepsilon_0c^2R_0}. \label{k3}
\end{equation}
\textit{+ If the charge is uniformly distributed in the interval $r$ with $R_e < r < R_0 $,} then by \eqref{vk3}
\begin{eqnarray}
&&M_{(e)ins}=\nonumber\\
&&\frac{Q^2\left[ \frac{1}{5}\left(R_0^5-R_e^5 \right)-R_e^3\left(R_0^2-R_e^2 \right)-R_e^6\left( \frac{1}{R_0}-\frac{1}{R_e}\right) \right]}{8\pi\varepsilon_0c^2\left(R_0^3-R_e^3 \right)^2 },\nonumber \label{k5fr}
\end{eqnarray}
\begin{eqnarray}
&&M_f =\nonumber\\  
&&\frac{Q^2\left[ \frac{1}{5}\left(R_0^5-R_e^5 \right)-R_e^3\left(R_0^2-R_e^2 \right)-R_e^6\left( \frac{1}{R_0}-\frac{1}{R_e}\right) \right]}{8\pi\varepsilon_0c^2\left(R_0^3-R_e^3 \right)^2 }\nonumber\\
&&+ ~M + \frac{Q^2}{8\pi\varepsilon_0c^2R_0}. \label{k5}
\end{eqnarray}\\

Notice that: The $g_{\mu\nu}$ tensor metrics in \eqref{k1} are outside the charged gravitational source and time-independent (stationary state). The $ g_{\mu\nu} $ tensor metrics inside the charged gravitational source which depends on time, we can see \eqref{Aa8} and \eqref{Aa9} in appendix.\\

It is easy to see that the metric \eqref{k1} has two singular solutions ($g_{00}=0 $, $ g_{11}=\infty $) that are $ r_{sq1}=\frac{2GM_f}{c^2}\left(\frac{1}{2}-\frac{1}{2}\sqrt{1-\frac{Q^2}{4\pi \varepsilon_0GM^2_f}} \right) $ and
\begin{equation}
r_{sq}=\frac{2GM_f}{c^2}\left(\frac{1}{2}+\frac{1}{2}\sqrt{1-\frac{Q^2}{4\pi \varepsilon_0GM^2_f}} \right). \label{k10}
\end{equation}
From \eqref{k10} we see, in order to have the expression under the square root  not negative then
\begin{equation}
\mid Q \mid \leq \sqrt{4\pi \varepsilon_0 G}M_f. \label{k6}
\end{equation}
When $Q=0 $ then $r_{sq1}=0 $ and $r_{sq}=r_s$ with
\begin{equation}
r_s=\frac{2GM}{c^2}. \label{rs}
\end{equation}
Thus, when the gravitational source is uncharged, the event horizon of the black hole $r_{sq}$ will return to the event horizon of the Schwarzschild black hole $r_s$. Also easy to see that $ r_{sq1} \leq r_{sq} $ and $ r_{s} \leq r_{sq} $. Below we will show that the value $r_{sq1}$ will be eliminated. It means that the charged gravitational source also has only one  event horizon of the black hole, it is $r_{sq}$.\\

\begin{figure}[h]
\begin{center}
\includegraphics[scale=0.62]{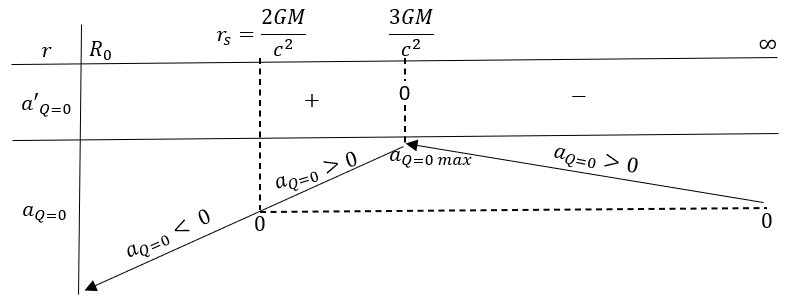}
\caption{\label{fig1}\textit{The variation table of $ a_{Q=0} $ with respect to $ r $}.}
\end{center}
\end{figure}

\begin{figure}[h]
\begin{center}
\includegraphics[scale=0.62]{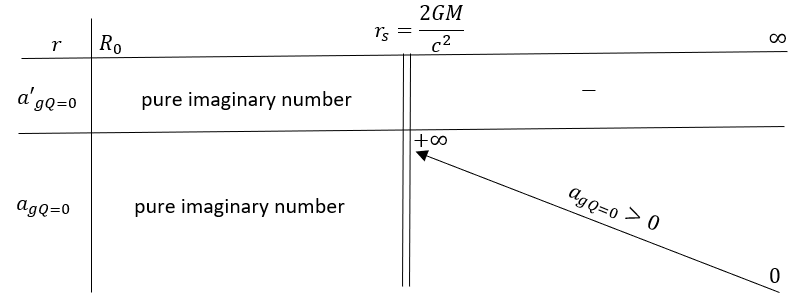}
\caption{\label{fig2}\textit{The variation table of $ a_{gQ=0} $ with respect to $ r $}.}
\end{center}
\end{figure}

Now let's consider the gravitational acceleration: For simplicity, we consider a neutral particle moving from the outside to the center of the field along an orbit of radius $r$. The equation of the motion of a particle in curved space-time is (\cite{L2})
\begin{equation}
\frac{d^2 x^\mu}{ds^2}+ \Gamma^\mu_{\alpha\beta}\frac{dx^\alpha}{ds}\frac{dx^\beta}{ds}=0. \label{cd1}
\end{equation}
We have $ d\theta=0 $, $ d\varphi=0 $ and notice that $ \Gamma^\mu_{\alpha\beta} $ in the spherically symmetric field, the equation \eqref{cd1} becomes
\begin{equation}
\frac{d^2 r}{ds^2} + \Gamma^1_{00}\left(\frac{dx^0}{ds}\right)^2 + \Gamma^1_{11}\left( \frac{dr}{ds}\right)^2=0. \label{cd2}
\end{equation}
Where $ x^0=ct $ and $ ds^2=g_{00}{dx^0}^2+g_{11}dr^2$. Since we want to investigate the particle's acceleration at a point, at that point we can assume that the particle's velocity is zero, $dr=0$. The equation \eqref{cd2} becomes 
\begin{equation}
\frac{d^2r}{dt^2}=\frac{c^2}{2}\frac{g'_{00}}{g_{11}}, \label{cd3}
\end{equation}
(notice $ \Gamma^1_{00}=-\dfrac{1}{2}\dfrac{g'_{00}}{g_{11}} $). Set $ a=-\frac{d^2r}{dt^2}$ (the sign ``$-$'' means that, by convention, when the particle goes in the direction from the outside to the center of the field, $ a > 0$). Then combining with metric \eqref{k1} we get
\begin{equation}
a=\left( \frac{GM_f}{r^2}-\frac{{G}Q^2}{4\pi\varepsilon_0c^2r^3}\right)\left( 1-\frac{2GM_f}{c^2r}+\frac{{G}Q^2}{4\pi\varepsilon_0c^4r^2}\right). \label{cd4}
\end{equation}
In the case of an uncharged gravitational source $Q=0 $, \eqref{cd4} becomes
\begin{equation}
a_{Q=0}=\frac{GM}{r^2}\left( 1-\frac{2GM}{c^2r}\right). \label{cd5}
\end{equation}
it follows that $ a'_{Q=0}=\frac{da_{Q=0}}{dr}=-\frac{2GM}{r^3} \left(1-\frac{3GM}{c^2r} \right) $. We have the variation table of $ a_{Q=0} $ with respect to $ r $ as shown in Fig. \ref{fig1}. 
From the Fig. \ref{fig1} we see: from the outside to the center of the field, $ a_{Q=0} $ increases to the maximum value $ a_{Q=0~ max}=\frac{c^4}{27GM} $ at the point $r=\frac{3GM}{c^2} $ (not at the Schwarzschild radius $r_s$). When passing the point $ r=\frac{3GM}{c^2} $, the quantity $ a_{Q=0} $ begins to decrease but still is positive  ($ a_{Q=0} > 0 $). After going through the Schwarzschild radius, $ a_{Q=0} $ starts to change sign ($ a_{Q=0} < 0 $). Does the event $ a_{Q=0} $ takes a negative value when $ r < r_s $ indicate that the gravitational field generates a force of repulsion in this region? This conclusion is incorrect, indeed. We notice that the quantity $ a=-\frac{d^2r}{dt^2} $ does not carry the meaning of  the acceleration due to gravity, because the quantity $dr $ is not a real distance and $dt$ is also not real time. We have to find the gravitational acceleration $ a_g $ in terms of real distance $ dl $ and real time $ d\tau $ (see the formulas \eqref{tau1} and \eqref{ds}). The gravitational acceleration $ a_g $ is
\begin{equation}
a_g=\frac{d^2l}{d\tau^2}, \label{gt1}
\end{equation}
to distinguish it from  $a= \frac{d^2r}{dt^2} $. Both of the articles \cite{Gron} and \cite{Asghar} misdefined the mathematical expression of the gravitational acceleration, they defined the gravitational acceleration as $ \frac{d^2r}{d\tau^2} $. To have the correct expression, $ d^2r $ must be replaced by $ d^2l $.  In a spherically symmetric field, from \eqref{ds} we have (notice $ d\theta=0 $ and $ d\varphi=0 $) $ dl^2= -g_{11}dr^2 $,
or
\begin{equation}
dl=-\sqrt{-g_{11}}dr, \label{gt3}
\end{equation}
(because the particle moving in the direction from outside to the center of the field $dr < 0 $.) Combining \eqref{gt3}, \eqref{gt1} and \eqref{tau1} we obtain (notice that we are considering the velocity of the particle is zero at the point, $dr=0 $)
\begin{equation}
a_g= -\frac{\sqrt{-g_{11}}}{g_{00}}\frac{d^2r}{dt^2}. \label{gt5}
\end{equation}
Combining \eqref{gt5} and \eqref{cd3}, we have
\begin{equation}
a_g=\frac{c^2}{2}\frac{g'_{00}}{g_{00}\sqrt{-g_{11}}}. \label{gt6}
\end{equation}
Using metric \eqref{k1}, the formula \eqref{gt6} becomes
\begin{equation}
a_g=\frac{\frac{GM_f}{r^2}\left(1-\frac{Q^2}{4\pi \varepsilon_0c^2M_f r} \right) }{\sqrt{ 1-\frac{2GM_f}{c^2r}+\frac{{G}Q^2}{4\pi\varepsilon_0c^4r^2}}}. \label{gt7}
\end{equation}
The formula \eqref{gt7} is also mentioned in \cite{Celerier:2017bny}. When the gravitational source is uncharged, $Q=0 $ then \eqref{gt7} becomes
\begin{equation}
a_{gQ=0}=\frac{\frac{GM}{r^2}}{\sqrt{ 1-\frac{2GM}{c^2r}}}. \label{gt8}
\end{equation}
It follows that $a'_{gQ=0}=\frac{d a_{gQ=0}}{dr}= \frac{-\frac{GM}{r^3}\left( 2-\frac{3GM}{c^2r}\right) }{\left(1-\frac{2GM}{c^2r} \right)\sqrt{1-\frac{2GM}{c^2r}} } $. From here we have the variation table of $ a_{gQ=0} $ with respect to $ r $ as shown in Fig. \ref{fig2}. 
From the Fig. \ref{fig2} we see that, in the direction from outside to the center of the gravitational field, the gravitational acceleration $ a_{gQ=0} $ increases, the gravitational field is always attractive $ a_{gQ=0} >0 $. When approaching the event horizon of the Schwarzschild black hole, the gravitational acceleration has the value of infinity ($ a_{gQ=0}=+\infty $), that is, the gravitational force has the value of infinity at the Schwarzschild radius. When $r < r_s $, the value of the gravitational acceleration will be a pure imaginary number. From \eqref{tau1} and \eqref{gt3} we see that the real time $ d\tau $ and the real distance $ dl $ are also a pure imaginary value. Thus, the gravitational field is always an attractive, $ a_{gQ=0} \geq 0$.\\
The pure imaginary values of the physical quantities just mentioned give us the following statements. The first possibility is that, the value of the Schwarzschild radius is the limit; the radius of the gravitational source cannot be reduced to less than the Schwarzschild radius. The second possibility is that, if the radius of  the gravitational source  can shrink smaller than the Schwarzschild radius, then the space region $r$ is between $R_0$ and $r_s$, $R_0 < r < r_s$, physical quantities  $ d\tau $, $ dl $, $ a_g $ take imaginary values. The imaginary values of physical quantities need to be clarified.\\[2mm]
For the charged gravitational source: From \eqref{gt7} we see $ a_g > 0 $ $ \forall r > R_0 $. That is, the gravitational field is always attractive as well. We can easily prove that $ a_g > 0 $ as follows: $ a_g < 0 $ only if $ r < r_k $ (where $ r_k = \frac{Q^2}{4\pi \varepsilon_0c^2M_f}$) . On the other hand, using \eqref{k6} we can easily prove that $r_k < r_{sq} $ (where $r_{sq} $ is the radius of the black hole's event horizon in \eqref{k10} ). However, when $ r< r_k < r_{sq} $ then $a_g $ will be a pure imaginary value (because the expression inside the square root \eqref{gt7} is negative). Thus, just like the uncharged gravitational source, when the gravitational source is charged, the gravitational field is always attractive and the gravitational force reaches infinity at the black hole's event horizon radius $r_{sq } $ (because when $ r=r_{sq} $ the gravitational acceleration  $ a_g = +\infty $).\\
We see: when the radius of the gravitational source $R_0 $ shrinks to the value of $ r_{sq} $, a black hole will be formed (gravitational force will be infinitely large at the event horizon $r_{sq} $). If $R_0 $ continues to shrink to less than $ r_{sq} $ then the physical quantities like $ dl $, $ d\tau $, $ a_g $ will take on  pure imaginary values. This fact gives us the following two statements: The first possibility is that, the value of $r_{sq} $ is limited and that $R_0 $ cannot be reduced to less than $ r_{sq} $, $ R_0 \geq r_{ sq} $. If the first possibility is true then $r_{sq1} $ does not exist because $r_{sq1} < r_{sq} $ (the solution \eqref{k1} is only used outside the gravitational source with $ r\geq R_0 $). The second possibility is that, if $R_0 $ can shrink to less than $ r_{sq} $ then the physical quantities will take on pure imaginary values. Therefore, the physical meaning of these pure imaginary values need to be clarified.

\section{Solution of the perturbative-$f(R)$ theory in the spherically symmetric field of a charged gravitational source}
In this section, we will find the solutions of the  perturbative-$f(R)$ theory around Einstein's equation; the method is performed in the same way as in the article \cite{Ky:2018fer} but for a charged spherically symmetric gravitational source.\\
 
In the GR theory, ${L}_G = R$ is the gravitational field's Lagrangian, $L^{(me)}$ is the Lagrangian of matter-radiation, the Einstein equation obtained is \eqref{2}--\eqref{RT}. In the $f(R)$ theory, the gravitational field's Lagrangian is ${L}_G = f(R)$, the system's action is 
\begin{align*}
S &= S_G + S^{(me)}\\
  &=\frac{c^3}{16\pi{G}}\int{f(R){\sqrt{-g}} {d^4x}}+\frac{1}{c}\int{L^{(me)}{\sqrt{-g}}{d^4x}}, 
\end{align*}
the  equation of the $ f(R) $ theory obtained is  (\cite{Nojiri:2010wj, DeFelice:2010aj, thomas, Capozziello:2011et})
\begin{align}
&f'(R)R_{\mu\nu}-g_{\mu\nu}\square f'(R)+\nabla_{\mu}\nabla_{\nu} f'(R)-\frac{1}{2}f(R)g_{\mu\nu}\nonumber\\
&=-kT_{\mu\nu}. 
\label{6}
\end{align}
Where $ k=\dfrac{8\pi G}{c^4} $, $ \nabla_\mu $ is the covariant
derivative and $ \square =\nabla_\mu\nabla^\mu $. As said in the introduction section, in the  perturbative-$f(R)$ theory, we have
\begin{equation}
f(R) = R + \lambda h(R), \label{hr1}
\end{equation}
where $ \lambda $ is constant and $h(R)$ is a general function that depends on the function $f(R)$, with the perturbation conditions
\begin{equation}
\mid \lambda h(R) \mid \ll \mid R \mid. \label{hr2}
\end{equation}
Combining \eqref{hr1} with \eqref{6}, then in perturbation terms (terms containing $\lambda$) of the resulting equation, replacing $ R=kT $ and $ R^\mu_\nu= -k\left( T^{\mu}_{~\nu}
-\frac{1}{2}\delta^{\mu}_{~\nu}T\right)$ (this is the Einstein equation), we get the first-order perturbation equation around Einstein's equation is (\cite{Ky:2018fer})
\begin{align}
&R^{\mu}_{~\nu}-\frac{1}{2}\delta^{\mu}_{~\nu}R-\lambda k h'(kT)\left(T^{\mu}_{~\nu}
-\frac{1}{2}\delta^{\mu}_{~\nu}T\right)  -\frac{\lambda}{2}\delta^{\mu}_{~\nu}h(kT)
\nonumber \\
&-\lambda \delta^{\mu}_{~\nu}\square^E h'(kT)+\lambda \nabla^{\mu}\nabla_{\nu}^Eh'(kT)
=-kT^{\mu}_{~\nu}. \label{9}
\end{align}
Where $h'(kT)=\frac{\partial h(kT)}{\partial (kT)}$ 
and the superscript $E$ in the covariant differentiations means that the metric tensor  $g_{\mu\nu}$ is 
taken in the Einstein's equation solutions. As mentioned above
\begin{equation}
T_{\mu\nu}= T^{(m)}_{\mu\nu} + T^{(e)}_{\mu\nu}, \label{fr1}
\end{equation}
with $ T^{(m)}_{\mu\nu} $ is the energy-momentum tensor of matter, $ T^{(e)}_{\mu\nu} $ is the energy-momentum tensor of the electromagnetic field.
On the other hand, the trace $ T^{(e)}=T^{\mu(e)}_{~\mu}=0 $, so 
\begin{equation}
T=T^{(m)}. \label{fr2}
\end{equation}
Using \eqref{fr2} to consider the perturbation terms (terms containing $ \lambda $) on the left hand side of the equation \eqref{9} we see that, although the charge term $Q $ does not contribute to $T$ (because $T=T^{(m)} $) but $Q $ still  contribute to components $ \nabla^{E}_\mu $, $ \square^E $ and $T^\mu_{~\nu} $.\\
Substituting the  metric tensor \eqref{11} into \eqref{9}, we get the result (see details in \cite{Ky:2018fer})
\begin{align}
&v(r,t)=\nonumber\\
&-\mbox{ln}\left\lbrace 1-\frac{1}{r}\int_0^r\left[ kT^0_{~0}
-\frac{\lambda}{2}h(kT)-\lambda \nabla^i \nabla_i^E h'(kT)\right.\right.\nonumber \\
&\left.\left. -\lambda k\left(T^0_{~0}
-\frac{T}{2}\right) h'(kT)\right] r'^2dr'\right\rbrace. \label{18}
\end{align}
Where $ T^0_{~0}=T^0_{~0}(r',t) $ and  $ T=T(r',t) $,
 \begin{align}
  &\nabla^i\nabla_i^Eh'(kT)=-\frac{1}{2c^2}\frac{g^{00}_E}{g^{11}_E}\frac{\partial g^{11}_E}
  {\partial t}\frac{\partial h'(kT)}{\partial t}+g^{11}_E\frac{\partial^2h'(kT)}{\partial r^2}\nonumber \\ 
&+\left( \frac{2}{r}g^{11}_E+\frac{1}{2}\frac{\partial g^{11}_E}{\partial r}\right) 
\frac{\partial h'(kT)}{\partial r}. \label{22}
 \end{align}
Here, as said before, the subscript $E$ indicates the Einstein limit. Using \eqref{fr2} we re-write \eqref{18} and \eqref{22} as
\begin{align}
&v(r,t)=\nonumber\\
&-\mbox{ln}\left\lbrace 1-\frac{1}{r}\int_0^r\left[ kT^0_{~0}
-\frac{\lambda}{2}h(kT^{(m)})-\lambda \nabla^i \nabla_i^E h'(kT^{(m)})\right.\right.\nonumber\\
&\left.\left. -\lambda k\left(T^0_{~0}
-\frac{T^{(m)}}{2}\right) h'(kT^{(m)})\right] r'^2dr'\right\rbrace \label{18fr}
\end{align}
and
 \begin{align}
  &\nabla^i\nabla_i^Eh'(kT^{(m)})=\nonumber\\
  &-\frac{1}{2c^2}\frac{g^{00}_E}{g^{11}_E}\frac{\partial g^{11}_E}
  {\partial t}\frac{\partial h'(kT^{(m)})}{\partial t}+g^{11}_E\frac{\partial^2h'(kT^{(m)})}{\partial r^2}\nonumber \\ 
&+\left( \frac{2}{r}g^{11}_E+\frac{1}{2}\frac{\partial g^{11}_E}{\partial r}\right) 
\frac{\partial h'(kT^{(m)})}{\partial r}. \label{22fr}
 \end{align}
\textbf{\subsection{Solutions outside the gravitational source}}
In this section, we find the solutions outside  the charged gravitation source. In perturbation terms (containing $\lambda$) we can neglect the  pressure (this can be seen as ignoring small amounts of higher order).
It means that the tensor $T\simeq T^0_{~0}$, \eqref{18fr} and \eqref{22fr} become
\begin{align}
v(r,t)=&-\mbox{ln}\left\lbrace 1-\frac{1}{r}\int_0^r\left[ kT^0_{~0}
-\frac{\lambda}{2}h(kT^{0(m)}_{~0})\right.\right.\nonumber\\
&\left.\left.-\lambda \nabla^i \nabla_i^E h'(kT^{0(m)}_{~0})\right.\right.\nonumber \\
&\left.\left. -\lambda k\left(T^0_{~0}
-\frac{T^{0(m)}_{~0}}{2}\right) h'(kT^{0(m)}_{~0})\right] r'^2dr'\right\rbrace \label{23fr}
\end{align}
and
 \begin{align}
  &\nabla^i\nabla_i^Eh'(kT^{0(m)}_{~0})=\nonumber\\
  &-\frac{1}{2c^2}\frac{g^{00}_E}{g^{11}_E}\frac{\partial g^{11}_E}
  {\partial t}\frac{\partial h'(kT^{0(m)}_{~0})}{\partial t}+g^{11}_E\frac{\partial^2h'(kT^{0(m)}_{~0})}{\partial r^2}\nonumber \\ 
&+\left( \frac{2}{r}g^{11}_E+\frac{1}{2}\frac{\partial g^{11}_E}{\partial r}\right) 
\frac{\partial h'(kT^{0(m)}_{~0})}{\partial r}. \label{24fr}
 \end{align}
Notice that: $ T=T^{(m)}+ T^{(e)} = T^{(m)} $ (because $ T^{(e)}=0 $). Outside the gravitational source $T^{\mu(m)}_{~\nu} =0$ and $T^{0(e)}_{~0}=T^{1(e)}_{ ~1} $ (see \eqref{me13}) so $ T^0_{~0}=T^1_{~1} $. That is,  outside the gravitational source, from the perturbative equation \eqref{9} we see that the equation \eqref{me2} is still true,
\begin{equation}
u(r, t) = -v(r, t). \label{me2fr}
\end{equation}
Notice that in the  $f(R)$ theory, the equation \eqref{me4} is incorrect, so the metric of a spherically symmetric field outside the gravitational source can still depend on time (see \cite{ Ky:2018fer}).\\
In \eqref{23fr}, to calculate the first integral term, doing the same calculations which from \eqref{C2a} to get \eqref{k1},  we obtain a perturbative solution outside the gravitational source 
\begin{align}
&g_{00}(r,t)= 1-\frac{kc^2M_f}{4\pi r}+\dfrac{kQ^2}{32\pi^2\varepsilon_0r^2}\nonumber\\
&+\frac{\lambda}{r}\int_0^r\left[\frac{1}{2}h(kT^{0(m)}_{~0})+ \nabla^i \nabla_i^E h'(kT^{0(m)}_{~0})\right.\nonumber \\
&\left. + k\left(T^0_{~0}
-\frac{T^{0(m)}_{~0}}{2}\right) h'(kT^{0(m)}_{~0})\right] r'^2dr',     \label{31g}
\end{align}
\begin{align}
&g_{11}(r,t)=-\left\lbrace 1-\frac{kc^2M_f}{4\pi r}+\dfrac{kQ^2}{32\pi^2\varepsilon_0r^2}\right.\nonumber\\
&+\frac{\lambda}{r}\int_0^r\left[\frac{1}{2}h(kT^{0(m)}_{~0})+ \nabla^i \nabla_i^E h'(kT^{0(m)}_{~0})\right.\nonumber \\
&\left.\left. + k\left(T^0_{~0}
-\frac{T^{0(m)}_{~0}}{2}\right) h'(kT^{0(m)}_{~0})\right] r'^2dr'\right\rbrace^{-1} , \label{32g}
\end{align}
\begin{align}
g_{22}=-r^2, \label{33g}
\end{align}
\begin{align}
g_{33}=-r^2sin^2\theta.\label{34g}
\end{align}
Where $ k=\frac{8\pi G}{c^4} $, $ T^0_{~0}=T^0_{~0}(r',t) $, $ h'(kT^0_{~0})
=\frac{\partial h(kT^0_{~0})}{\partial (kT^0_{~0})} $, 
$ \nabla^i \nabla_i^E h'(kT^0_{~0}) $ 
calculated in \eqref{24fr} and $ M_f $ calculated in \eqref{k2}---\eqref{k5}. Using
\begin{equation}
T^0_{~0}= T^{0(e)}_{~0}+T^{0(m)}_{~0}, \label{Tem}
\end{equation}
we have
\begin{align}
&g_{00}(r,t)= 1-\frac{kc^2M_f}{4\pi r}+\dfrac{kQ^2}{32\pi^2\varepsilon_0r^2}\nonumber\\
&+\frac{\lambda}{r}\int_0^r\left[\frac{1}{2}h(kT^{0(m)}_{~0})+ \nabla^i \nabla_i^E h'(kT^{0(m)}_{~0})\right.\nonumber \\
&\left. + k\left(T^{0(e)}_{~0}
+\frac{T^{0(m)}_{~0}}{2}\right) h'(kT^{0(m)}_{~0})\right] r'^2dr',     \label{31}
\end{align}
\begin{align}
&g_{11}(r,t)=-\left\lbrace 1-\frac{kc^2M_f}{4\pi r}+\dfrac{kQ^2}{32\pi^2\varepsilon_0r^2}\right.\nonumber\\
&+\frac{\lambda}{r}\int_0^r\left[\frac{1}{2}h(kT^{0(m)}_{~0})+ \nabla^i \nabla_i^E h'(kT^{0(m)}_{~0})\right.\nonumber \\
&\left.\left. + k\left(T^{0(e)}_{~0}
+\frac{T^{0(m)}_{~0}}{2}\right) h'(kT^{0(m)}_{~0})\right] r'^2dr'\right\rbrace ^{-1}, \label{32}
\end{align}
\begin{align}
g_{22}=-r^2, \label{33}
\end{align}
\begin{align}
g_{33}=-r^2sin^2\theta. \label{34}
\end{align}
When considering the $g_{\mu\nu}$ metric that is far from the gravitational source, in perturbation terms (containing $\lambda$) we can consider $ T^0_{~0} $ depends only on the time $ t $ (the density of the gravitational source can be considered homogeneous, this can be seen as ignoring small amounts of higher order in the perturbation terms),
that means the last two terms of \eqref{24fr} are vanished, 
%\eqref{22} and we have
\begin{align}
\nabla^i \nabla_i^E h'(kT^{0(m)}_{~0})=-\frac{1}{2c^2}\frac{g^{00}_E}{g^{11}_E}
\frac{\partial g^{11}_E}{\partial t}\frac{\partial h'(kT^{0(m)}_{~0})}{\partial t}. \label{35}
\end{align}
It follows that  [notice that outside the gravitational source $ r \geq R_0(t) $ (with $R_0(t)$ is the radius of the gravitational source), $ T^{0(m)}_{~0}=0 $ and $ h(kT^{0(m)}_{~0})=h(0)=\mbox{constant} $]
 \begin{eqnarray}
&&\int_0^{r}\nabla^i \nabla_i^E h'(kT^{0(m)}_{~0})r'^2dr'\nonumber\\
&&=\int_0^{R_0(t)}\nabla^i \nabla_i^E h'(kT^{0(m)}_{~0})r'^2dr'\nonumber \\
&&\approx h''(kT^{0(m)}_{~0})\left[ \frac{\partial}{\partial t}
\frac{M}{[R_0(t)]^3}\right]^2 \alpha (t), \label{36}
\end{eqnarray}
(see \eqref{36} more details in the appendix). Where, 
\begin{align}
\alpha (t)=&\frac{3k^2c^2R_0(t)}{256\pi^2\vartheta(t)[\xi (t)]^4}\left\lbrace \frac{3}{\xi(t)R_0(t)}
\arcsin[\xi (t) R_0(t)]\right.\nonumber \\
 &\left. -\left( 3+2[\xi(t)R_0(t)]^2\right) \sqrt{1-[\xi(t)R_0(t)]^2}\right\rbrace, \label{37}
\end{align}
with  
\begin{align}
\xi^2 (t)=\frac{kMc^2}{4\pi [R_0(t)]^3}, \label{37a}
\end{align}
and
\begin{equation}
\vartheta(t)=\left( 1-\frac{2GM_f}{c^2R_o(t)}+\frac{{G}Q^2}{4\pi\varepsilon_0c^4[R_0(t)]^2}\right) \sqrt{1-\frac{2GM}{c^2R_o(t)}}. \label{37aq}
\end{equation}
Substituting  \eqref{36} into  \eqref{31} -- \eqref{34} we find a solution at a 
distant point from the gravitational source: 
\begin{align}
g_{00}(r,t)=&~ 1-\frac{kc^2M_f}{4\pi r}+\dfrac{kQ^2}{32\pi^2\varepsilon_0r^2}+\frac{\lambda}{r}\int_0^r\left[\frac{1}{2}h(kT^{0(m)}_{~0})\right.\nonumber \\
&\left. + k\left(T^{0(e)}_{~0}
+\frac{T^{0(m)}_{~0}}{2}\right) h'(kT^{0(m)}_{~0})\right] r'^2dr'\nonumber \\
&+\frac{\lambda h''(kT^{0(m)}_{~0})}{r}\left[ \frac{\partial}{\partial t}
\frac{M}{[R_0(t)]^3}\right]^2 ~\alpha (t), \label{38}
\end{align}
\begin{align}
&g_{11}(r,t)=\nonumber\\
&-\left\lbrace 1-\frac{kc^2M_f}{4\pi r}+\dfrac{kQ^2}{32\pi^2\varepsilon_0r^2}+\frac{\lambda}{r}\int_0^r\left[\frac{1}{2}h(kT^{0(m)}_{~0})\right.\right.\nonumber \\
&~~~~~\left. + k\left(T^{0(e)}_{~0}
+\frac{T^{0(m)}_{~0}}{2}\right) h'(kT^{0(m)}_{~0})\right] r'^2dr'\nonumber \\
&~~~~~\left.+\frac{\lambda h''(kT^{0(m)}_{~0})}{r}\left[ \frac{\partial}{\partial t}
\frac{M}{[R_0(t)]^3}\right]^2 ~\alpha (t)\right\rbrace ^{-1}, \label{39}
\end{align}
\begin{align}
g_{22}=-r^2, \label{40}
\end{align}
\begin{align}
g_{33}=-r^2\sin^2\theta. \label{41}
\end{align}
Where 
\begin{equation}
T^{0(m)}_{~0}(t) = \frac{Mc^2}{\frac{4}{3}\pi [R_0(t)]^3}
\label{24}
\end{equation}
 is used for both inside and outside the integral, and 
$ h''(kT^0_{~0})=\frac{\partial^2h(kT^0_{~0})}{\partial(kT^0_{~0})^2} $. Although $ T^0_{~0} $ does not depend on $r$ (only on $t$), one must be careful when bringing $ \frac{1}{2}h(kT^{0(m)}_{~0}) + k\left(T^{0(e)}_{~0}+\frac{T^{0(m)}_{~0}}{2}\right) h'(kT^{0(m)}_{~0}) $ out of the integral.  
If $ \frac{1}{2}h(kT^{0(m)}_{~0}) + k\left(T^{0(e)}_{~0}+\frac{T^{0(m)}_{~0}}{2}\right) h'(kT^{0(m)}_{~0})$ is zero in vacuum, then the integral is performed 
inside the gravitational source. But sometimes\\ $ \frac{1}{2}h(kT^{0(m)}_{~0}) + k\left(T^{0(e)}_{~0}+\frac{T^{0(m)}_{~0}}{2}\right) h'(kT^{0(m)}_{~0}) $ is non-zero in vacuum, see some examples below.\\

The \eqref{38}--\eqref{41} solution is general, which can be used for any model.  Now, we will consider some special model.\\

\subsubsection{The case $ f(R)=R-2\lambda$ (model I)}
 This model is the GR theory with the cosmological constant $ \Lambda=\lambda $. In this model, since $ h(R)=-2 $, $ h(kT^0_{~0})=-2 $ and $ h'(kT^0_{~0})=0 $.
 Using \eqref{38}--\eqref{41}, it is easy to see
\begin{align}
g_{00}(r,t)=1-\frac{kc^2M_f}{4\pi r}+\dfrac{kQ^2}{32\pi^2\varepsilon_0r^2}-\frac{\lambda r^2}{3}, \label{42}
\end{align}
\begin{align}
g_{11}(r,t)=\frac{-1}{1-\frac{kc^2M_f}{4\pi r}+\dfrac{kQ^2}{32\pi^2\varepsilon_0r^2}-\frac{\lambda r^2}{3}}, \label{43}
\end{align}
\begin{align}
g_{22}=-r^2, \label{44}
\end{align}
\begin{align}
g_{33}=-r^2sin^2\theta. \label{45}
\end{align}
This solution is the exact solution of the GR theory with the cosmological constant  $\lambda$.\\

\subsubsection{The case $ f(R)=R+\lambda R^b $ ($b>0$), (model II)}
The $f(R)$ theory must satisfy two conditions $\frac{df(R)}{dR}  > 0$ and $\frac{d^2f(R)}{dR^2} >0$. The first condition is for the effective gravitational constant to be positive (\cite{Jaime:2010kn}), the second condition is to stop the Dolgov-Kawasaki instability (\cite{Dolgov:2003px}), so we choose $b>0$.\\

In this model, since $ h(R)=R^b $, $h'(R)= bR^{b-1} $ and $ h''(R)=b(b-1)R^{b-2} $. The 
  \eqref{38} --  \eqref{41} formulas become
\begin{align}
&g_{00}(r, t)=1-\frac{kc^2M_f}{4\pi r}+\dfrac{kQ^2}{32\pi^2\varepsilon_0r^2}\nonumber\\
&+\frac{\lambda (b+1)k^b }{2r}\int^{R_o(t)}_0{[T^{0(m)}_{~0}]}^br'^2dr'\nonumber\\
&+\frac{\lambda bk^b}{r}\int_0^{R_0(t)}T^{0(e)}_{~0}[T^{0(m)}_{~0}]^{b-1}r'^2dr'
\nonumber \\
&+\frac{\lambda}{r}b(b-1)k^{b-2}[T^{0(m)}_{~0}]^{b-2}\left[ \frac{\partial}
{\partial t}\frac{M}{[R_o(t)]^3}\right]^2 ~\alpha (t), \label{46}
\end{align}
\begin{align}
&g_{11}(r, t)=-\left\lbrace 1-\frac{kc^2M_f}{4\pi r}+\dfrac{kQ^2}{32\pi^2\varepsilon_0r^2}\right.\nonumber\\
&+\frac{\lambda (b+1)k^b }{2r}\int^{R_o(t)}_0{[T^{0(m)}_{~0}]}^br'^2dr'\nonumber\\
&+\frac{\lambda bk^b}{r}\int_0^{R_0(t)}T^{0(e)}_{~0}[T^{0(m)}_{~0}]^{b-1}r'^2dr'
\nonumber \\
&\left.+\frac{\lambda}{r}b(b-1)k^{b-2}[T^{0(m)}_{~0}]^{b-2}\left[ \frac{\partial}
{\partial t}\frac{M}{[R_o(t)]^3}\right]^2 ~\alpha (t)\right\rbrace ^{-1}, \label{47}
\end{align}
\begin{align}
g_{22}=-r^2, \label{48}
\end{align}
\begin{align}
g_{33}=-r^2sin^2\theta. \label{49}
\end{align}
Further, using \eqref{24}, \eqref{meg1} and \eqref{k2fr} we obtain
\begin{align}
&g_{00}(r,t)=1-\frac{kc^2M_{ff}(t)}{4\pi r}+\dfrac{kQ^2}{32\pi^2\varepsilon_0r^2}, \label{50}\\
&g_{11}(r,t)=\frac{-1}{1-\frac{kc^2M_{ff}(t)}{4\pi r}+\dfrac{kQ^2}{32\pi^2\varepsilon_0r^2}}, \label{51}\\
&g_{22}=-r^2, \label{52}\\
&g_{33}=-r^2\mbox{sin}^2\theta. \label{53}
\end{align}
Here
\begin{equation}
M_{ff}(t)=M_f-\lambda M_1(t)-\lambda M_2(t)-\lambda {M_3}(t), \label{mff}
\end{equation}
\begin{align}
&M_1(t)=\frac{4\pi}{kc^2}\frac{(b+1)c^{2b}(kM)^b }{3^{1-b}~2^{2b+1}~{\pi}^b[R_o(t)]^{3b-3}}, \label{53a}\\
&M_2(t)=\frac{4\pi}{kc^2}\frac{b(b-1)c^{2b-4}(3kM)^{b-2}\left[ 
\frac{\partial}{\partial t}\frac{M}{[R_o(t)]^3}\right]^2 \alpha (t)}{(4\pi)^{b-2}[R_o(t)]^{3b-6}},\label{53b}
\end{align}
\begin{equation}
{M_3}(t)=\dfrac{bk^{b-1}M^{b-1}c^{2(b-1)}3^{b-1}}{(4\pi)^{b-1}[R_0(t)]^{3(b-1)}}M_{(e)ins}(t), \label{53bfr}
\end{equation}
where $ M_f$ and $ M_{(e)ins}(t) $ calculated in \eqref{k2}---\eqref{k5}. From \eqref{hr2}, the perturbation condition for this model  is (take an approximation of $ R\approx kT $)
\begin{equation}
\mid\lambda\mid ~\ll ~\mid(kT)^{1-b}\mid. \label{dk1}
\end{equation}
The condition \eqref{dk1} means that in \eqref{mff} we will have
\begin{equation}
\mid\lambda M_1(t)+\lambda M_2(t)+\lambda {M_3}(t)\mid~\ll~\mid M_f\mid. \label{dkbs}
\end{equation}
Notice that in the \eqref{dk1}  does not mean  $ \lambda $  small, it is large or small depending on the $ b $ as long as the perturbation condition is holded.
This perturbation method works in the presence of gravitational source $ T_{\mu\nu} $ (unlike the other works that only work with the vacuum).  For the model 
\begin{equation}
f(R)=R+\lambda R^2, \label{R2}
\end{equation}
($ b=2 $), with an uncharged gravitational source, the \eqref{dk1} perturbation condition is satisfied (for the solar system) if $ \mid\lambda\mid \ll 0.380053 \times 10^{23} m^2$. To correct the GR theory for Mercury's precession to be more consistent with the observations, we take $ \lambda=0.1511677 \times 10^{18} m^2 $ (see details in \cite{Ky:2018fer}).  The perturbative-$f(R)$ theory, it is a small correction to Einstein's theory in Lagrangian. It is not the same as the approximate solution method of the field equation when the space-time curvature is small.
Now consider this model for black holes: Consider the black hole Sgr A* (the black hole at the center of the Milky Way) with mass $ M = 4.31\times 10^6 M_{\odot}=8.57 \times 10 ^{36} kg $ ($ M_{\odot} $ is the mass of the Sun) and radius $ R_0 = 22 \times 10^9 m $ (\cite{Gillessen:2008qv}). From these data we see that  the perturbation condition \eqref{dk1} for this black hole would be $\mid\lambda\mid \ll 0.2789 \times 10^{21} m^2$ . Using perturbation solutions to study effects when the star S2 orbits Sgr A* can be found in the conclusions section of \cite{Ky:2018fer}. Here we introduce another effect, from the perturbation solutions \eqref{50}--\eqref{53} we can easily see that the event horizon of the black hole is 
\begin{equation}
r_{sqf}=\frac{2GM_{ff}}{c^2}\left(\frac{1}{2}+\frac{1}{2}\sqrt{1-\frac{Q^2}{4\pi \varepsilon_0GM^2_{ff}}} \right). \label{rsqf}
\end{equation}
Comparing \eqref{rsqf} with \eqref{k10} we see that the  event horizon of the perturbative-$f(R)$ theory corrected  the GR theory by  substituting $M_{ff} $ for $ M_f $. Notice that, when the field is static then $M_2 =0 $, only $M_1$ and $M_3$ left. If the field is static and the gravitational source is uncharged, there remains only the term $ M_1 $. For Sgr A* we can easily calculate $ M_1 = 4.609\times 10^{16}kgm^{-2}$. With the perturbation condition of Sgr A* as mentioned above: If taking $ \lambda = 0.1511677 \times 10^{18} m^2 $ (see above), then $ \lambda M_1= 6.967\times 10^{ 33} kg$, the correction for mass $M$ (we get the effective mass $M_{ff}$) is about $ \lambda M_1/M=8.13\times 10^{-4} = 0.0813\% $ (notice that, depends on the intended application, if we take a different $ \lambda $ value, we will get a different result). The event horizon of the black hole Sgr A* would be 
\begin{equation}
\frac{r_{sf(R)}}{r_{sE}} = \frac{M_{ff}}{M}=\frac{M-\lambda M_1}{M} = 0.999187, \label{rsf}
\end{equation}
where $ r_{sE} $ is the event horizon according to the Einstein theory, $ r_{sf(R)} $ is the event horizon according to the model $ f(R)=R + \lambda R^2 $.\\

Thus, the solution of this model corrected  the solution of Einstein's equation by replacing $M_f$ by $M_{ff}$. The  metric tensors outside the gravitational source are now  time-dependent if the gravitational source radius $R_0(t) $ depends on time (the radius of the gravitational source expands or shrinks). While in Einstein's equation, these metric tensors are always stationary (they do not depend on time). The  metric tensors are time-dependent, which can cause an interesting effect, that is, unlike in Einstein's equation. In the $f(R)$ theory, a spherically symmetric field can still emit gravitational waves. In particular, this perturbative-$f(R)$ theory has shown that when a collapsing star approaches its event horizon, it emits very strong gravitational waves \cite{Ky:2024lce}. However, it is easy to see that,  the formulas \eqref{me6}--\eqref{meg1} are still true. It means that,  a spherically symmetric field still does not emit electromagnetic waves.\\

Meaning of the correction terms $\lambda M_1(t)$, $\lambda M_2(t)$, $\lambda {M_3}(t)$ is that, they will produce gravitational wave radiation of a non-static spherically symmetric gravitational field. In addition, when the source of spherical symmetry is static it also gives corrections to Einstein's theory to make the physical quantities more precise. For instance, it will describe the precession of the planets more accurately (see details in \cite{Ky:2018fer, Ky:2019gbj, VanKy:2020xxj}).\\

\subsubsection{ The case $ f(R)=R-2\Lambda+\gamma R^b $ ($b>0$), (model III)}
In this model we have $\lambda h(R)= -2\Lambda+\gamma R^b$. Similar calculation, using \eqref{38} -- \eqref{41} we get the solution of this model as
\begin{align}
&g_{00}(r,t)=1-\frac{kc^2M_{ff}(t)}{4\pi r}+\dfrac{kQ^2}{32\pi^2\varepsilon_0r^2}-\frac{\Lambda r^2}{3}, \label{iii1}\\
&g_{11}(r,t)=\frac{-1}{1-\frac{kc^2M_{ff}(t)}{4\pi r}+\dfrac{kQ^2}{32\pi^2\varepsilon_0r^2}-\frac{\Lambda r^2}{3}}, \label{iii2}\\
&g_{22}=-r^2, \label{iii3}\\
&g_{33}=-r^2\mbox{sin}^2\theta. \label{iii4}
\end{align}
Where $M_{ff}(t) $ is the formulas \eqref{mff}-\eqref{53bfr} (with $ \lambda $ replaced by $ \gamma $).\\

In cosmology, the solution of the perturbative $f(R)$ theory can be seen in \cite{VanKy:2022itq}, which shows that the perturbative $f(R)$ theory solution can be viewed as a GR theory with the cosmological constant $ \Lambda $ replaced by the effective cosmological constant $ \Lambda_{eff}(\rho(t)) $. It can describe the evolution of the universe in both the inflationary epoch and current epoch (dark energy problem).\\

\subsubsection{Some other models:}
We will show the way to implement the theories $f(R)$ where the function $f(R)$ is not in the form of $f(R)=R+\lambda h(R) $ as model I, model II and model III. For example, consider the model $f(R)=R^{1+\varepsilon} $ with $ \mid\varepsilon\mid\ll 1 $. We can then do the following, writing $ f(R)=R^{1+\varepsilon}=R+ \left(R^{1+\varepsilon}-R \right) $,  we see $ \lambda h(R)= R^{1+\varepsilon}-R $. From here, using the general solutions \eqref{38}--\eqref{41} we will get the perturbation solutions of the model $f(R)=R^{1+\varepsilon} $.  From the general solution \eqref{38}--\eqref{41} it is easy to see that the metric of these $f(R)$ models  has the same form as \eqref{50}--\eqref{53},
\begin{align}
&g_{00}(r,t)=1-\frac{kc^2M_{ff}(t)}{4\pi r}+\dfrac{kQ^2}{32\pi^2\varepsilon_0r^2}, \label{som1}\\
&g_{11}(r,t)=\frac{-1}{1-\frac{kc^2M_{ff}(t)}{4\pi r}+\dfrac{kQ^2}{32\pi^2\varepsilon_0r^2}}, \label{som2}\\
&g_{22}=-r^2, \label{som3}\\
&g_{33}=-r^2\mbox{sin}^2\theta. \label{som4}
\end{align}
Where $M_{ff}(t) $ has  different forms depending on the form of the  $ f(R) $ function.\\

Nowadays, there are many authors looking for a model $f(R)$ that can be used for  the inflationary epoch, the dark energy problem, the dark matter problem, and satisfy physical phenomena in spherically symmetric fields (such as the solar system), but none of them is perfect. This perturbation method will contribute to finding a perfect $f(R)$ model in the future.

\textbf{\subsection{The general solution of the perturbative-$f(R)$ theory }}
As seen above, we found the solution of the perturbative-$f(R)$ theory outside the gravitational source. We now find the general solution that works both outside and inside of the gravitational source.
Since inside the gravitational source it do not have $ u(r,t)=-v(r,t) $, we can solve this problem in the following way: The same calculation as obtained the \eqref{22fr} formula, we get
\begin{align}
\square^E h'(kT^{(m)})-\nabla^1\nabla^E_1h'(kT^{(m)})=\beta (r,t), \label{58}
\end{align}
where,
\begin{align}
\beta (r, t)=&\frac{g^{00}_E}{c^2}\frac{\partial^2h'(kT^{(m)})}{\partial t^2}
+\frac{1}{2c^2}\frac{\partial g^{00}_E}{\partial t}\frac{\partial 
h'(kT^{(m)})}{\partial t}\nonumber \\
&+g^{11}_E\left( \frac{2}{r}-\frac{1}{2g^{00}_E}\frac{\partial 
g^{00}_E}{\partial r}\right) \frac{\partial h'(kT^{(m)})}{\partial r}. \label{59}
\end{align}
The $ E $ index means that the  metric tensors are taken from the GR theory 
(its values inside the gravitational source we can see the appendix). 
Substituting \eqref{58} into \eqref{9} we have 
\begin{align}
&R^{1}_{~1}-\frac{1}{2}R=-kT^{1}_{~1}+\frac{\lambda}{2}h(kT^{(m)})\nonumber \\
 &+\lambda k\left( T^{1}_{~1}-\frac{1}{2}T^{(m)}\right) h'(kT^{(m)})
+\lambda \beta (r,t), \label{60}
\end{align}
Using \eqref{29}, the \eqref{60} equation becomes
\begin{align}
&e^{-v(r,t)}\left[ \frac{u'(r,t)}{r}+\frac{1}{r^2}\right] 
-\frac{1}{r^2}=-kT^{1}_{~1}+\frac{\lambda}{2}h(kT^{(m)})\nonumber \\
&+\lambda k\left( T^{1}_{~1}-\frac{1}{2}T^{(m)}\right) h'(kT^{(m)})
+\lambda \beta (r,t). \label{61}
\end{align}
It follows that
\begin{align}
&u'(r,t)=r\left\lbrace e^{v(r,t)}\left[ \frac{1}{r^2}-kT^{1}_{~1}
+\frac{\lambda}{2}h(kT^{(m)})\right.\right.\nonumber \\
&\left.\left. +\lambda k\left( T^{1}_{~1}-\frac{1}{2}T^{(m)}\right) h'(kT^{(m)})+\lambda \beta (r,t)\right] 
-\frac{1}{r^2}\right\rbrace. \label{62}
\end{align}
Since $ e^{v(r,t)}=-g_{11}(r,t) $, we can rewrite the \eqref{62} equation as 
\begin{align}
&u'(r,t)=rg_{11}(r,t)\left[ -\frac{1}{r^2}+kT^{1}_{~1}-\frac{\lambda}{2}h(kT^{(m)}) \right.\nonumber \\
&\left.-\lambda k\left( T^{1}_{~1}-\frac{1}{2}T^{(m)}\right) h'(kT^{(m)})
-\lambda \beta (r,t)\right] -\frac{1}{r}. \label{63}
\end{align}
Since $g_{00}(r,t)\rightarrow 1 $ when $r\rightarrow \infty$, we obtain
\begin{align}
&u(r,t)=\int_\infty^r\left\lbrace r'g_{11}(r',t)\left[ -\frac{1}{r'^2}+kT^{1}_{~1}
-\frac{\lambda}{2}h(kT^{(m)})\right.\right.\nonumber \\
&\left.\left. -\lambda k\left( T^{1}_{~1}-\frac{1}{2}T^{(m)}\right) h'(kT^{(m)})-\lambda \beta (r',t)\right] 
-\frac{1}{r'}\right\rbrace dr'.  \label{64}
\end{align}
From \eqref{11}, \eqref{18fr} and \eqref{64} we obtain the general metric tensors of the perturbative-$f(R)$ theory, which can be used both inside and outside the gravitational source as 
\begin{align}
&g_{00}(r,t)=\exp\int_\infty^r\left\lbrace r'g_{11}(r',t)\left[ -\frac{1}{r'^2}
+kT^{1}_{~1}\right.\right.\nonumber\\
&~~~~~\left.\left.-\frac{\lambda}{2}h(kT^{(m)})-\lambda \beta (r',t) 
\right.\right.\nonumber \\
&~~~~~\left.\left.-\lambda k\left( T^{1}_{~1}-\frac{1}{2}T^{(m)}\right) h'(kT^{(m)})\right] 
-\frac{1}{r'}\right\rbrace dr',  \label{65}
\end{align}
\begin{align}
g_{11}(r,t)=&-\left\{1-\frac{1}{r}\int_0^r\left[ kT^0_{~0}-\frac{\lambda}{2}h(kT^{(m)})
\right.\right.\nonumber \\
&-\lambda k\left( T^0_{~0}-\frac{T^{(m)}}{2}\right) h'(kT^{(m)})\nonumber \\ 
&\left.\left.-\lambda \nabla^i \nabla_i^E h'(kT^{(m)})\right] r'^2dr'\right\}^{-1}, \label{66}
\end{align}
\begin{align}
g_{22}=-r^2, \label{67}
\end{align}
\begin{align}
g_{33}=-r^2\sin^2\theta, \label{68}
\end{align}
where $ \nabla^i \nabla_i^E h'(kT) $ and $ \beta(r',t) $ are   \eqref{22fr} and 
\eqref{59}, respectively.\\

\section{A embedding in the background of the FLRW cosmology}
The spherically symmetric metrics of a star (SSM) that we found above are based on a flat and non-expanding cosmological metric (FNEC). More precisely, we have to consider SSM in the background of a flat expanding cosmological metric (FEC). This section presents a method for embedding the Reissner-Nordström metric (a generalization of the Schwarzschild metric) in the background of the FLRW cosmology. As we know, many observations show that the metric of the universe today is flat, so here we will use the FLRW cosmological metric  which is in flat form (this is a metric with isotropic form).
\begin{align}
ds^2={dx^0}^2-a^2(t)\left[ dq^2+q^2d\theta^2+q^2\sin^2\theta{d\varphi}^2\right]. \label{n0}
\end{align}
Where $ a $ has the dimension of length, $ [a]=m $ and $ q $ has no dimension, $ [q]=1 $. If the scalar factor $ a $ depends on time $ a=a(t) $ we get an expanding universe (corresponding to SSM  embedded in the background of FEC). If $a = constant $ we will get an non-expanding universe (corresponding to SSM  embedded in the background of FNEC). To embed the SSM in the background of the cosmological metric, we must first perform a coordinate transformation to bring the SSM to the isotropic form.\\

We first consider a static Reissner-Nordström metric. In this case, we have $ \lambda M_2 =0$ but $ \lambda M_1=constant $ and $ \lambda M_3 =constant$, therefore  $ M_{ff}(t)=M_{ff} = constant $. The static SSM metric in Schwarzschild form is
\begin{align}
ds^2=g_{00}(r){dx^0}^2+g_{11}(r)dr^2-r^2d\theta^2-r^2\sin^2\theta{d\varphi}^2. \label{n1}
\end{align}
Its isotropic form is
\begin{align}
ds^2=A(q){dx^0}^2-B(q)\left[ dq^2+q^2d\theta^2+q^2\sin^2\theta{d\varphi}^2\right]. \label{n2}
\end{align}
Where $ [r]=m $, $ [g_{00}]=[g_{11}]=1 $; $ [q]=[A]=1 $, $ [B]=m^2 $. The metric \eqref{n1} can be converted to the form \eqref{n2} by coordinate transformation $ x^0=x^0 $, $ \theta=\theta $, $ \varphi=\varphi $, 
\begin{equation}
r=E(q). \label{n3}
\end{equation}
Where $E(q) $ is the function to find, $ [E]=m $. Replace \eqref{n3} into \eqref{n1} and then compare with \eqref{n2} we get
\begin{align}
&A(q)=g_{00}(E), \label{n4a}\\
&B(q)=-g_{11}(E)\left( \frac{dE}{dq}\right)^2,  \label{n4}\\
&B(q)q^2=E^2. \label{n4b}
\end{align}
From \eqref{n4} and \eqref{n4b} deduce
\begin{align}
\sqrt{-g_{11}(E)}\frac{dE}{E}=\frac{dq}{q}. \label{n5}
\end{align}
Follow \eqref{51} (or \eqref{som2}) with
\begin{equation}
g_{11}(E)=\frac{-1}{1-\frac{kc^2M_{ff}}{4\pi E}+\frac{kQ^2}{32\pi^2\varepsilon_0E^2}}, \label{n6}
\end{equation}
the equation \eqref{n5} becomes
\begin{equation}
\frac{dE}{\sqrt{\left( E-\overline{M}\right)^2+\overline{Q}^2-\overline{M}^2}}=\frac{dq}{q}. \label{n7}
\end{equation}
Where
\begin{align}
&\overline{M}=\frac{kc^2M_{ff}}{8\pi}, \label{n8}\\
&\overline{Q}^2=\frac{kQ^2}{32\pi^2\varepsilon_0}. \label{n9}
\end{align}
We can see $ [\overline{M}]=[\overline{Q}]=m $.  The general solution of the \eqref{n7} equation  is
\begin{equation}
E-\overline{M} + \sqrt{\left(E-\overline{M} \right)^2 + \overline{Q}^2-\overline{M}^2}=qC, \label{n10}
\end{equation}
where $C$ is any constant, $ [C]=m $. From here we have the function $E(q) $ is
\begin{equation}
E(q)=\frac{qC}{2}\left( 1+ \frac{\overline{M}}{qC}\right)^2-\frac{\overline{Q}^2}{2qC}. \label{n11}
\end{equation}
Combining \eqref{n4a} and \eqref{n5} we have
\begin{equation}
A(q)=g_{00}(E)=\frac{-1}{g_{11}(E)}=\left( \frac{q}{E}\frac{dE}{dq}\right)^2. \label{nn1}
\end{equation}
Substituting \eqref{n11} into \eqref{nn1} and \eqref{n4b} we get
\begin{equation}
A(q)=\left[ \frac{ 1-\frac{\overline{M}^2}{q^2C^2}+\frac{\overline{Q}^2}{q^2C^2}}{\left(1+\frac{\overline{M}}{qC} \right)^2-\frac{\overline{Q}^2}{q^2C^2}}\right]^2, \label{n12}
\end{equation}
\begin{equation}
B(q)=\left[ \frac{C}{2}\left(1+\frac{\overline{M}}{qC} \right)^2-\frac{\overline{Q}^2}{2q^2C}\right]^2. \label{n13}
\end{equation}
Thus, to convert a metric of the Schwarzschild form \eqref{n1} to the isotropic form \eqref{n2} we must use the coordinate transformation \eqref{n11}, $ r=E(q) $. Now, set $C=2a $, where $ a $ is a constant, $ [a]=m $, then the \eqref{n2} isotropic form is
\begin{align}
ds^2&=\left[ \frac{ 1-\frac{\overline{M}^2}{4q^2a^2}+\frac{\overline{Q}^2}{4q^2a^2}}{\left(1+\frac{\overline{M}}{2qa} \right)^2-\frac{\overline{Q}^2}{4q^2a^2}}\right]^2{dx^0}^2 \nonumber\\
&-a^2\left[\left(1+\frac{\overline{M}}{2qa} \right)^2-\frac{\overline{Q}^2}{4q^2a^2}\right]^2\nonumber\\
&\times\left[ dq^2+q^2d\theta^2+q^2\sin^2\theta{d\varphi}^2\right]. \label{n14}
\end{align}
From here, we see that if $M$ and $Q$ are zero then there will be only the background of FNEC left in the metric \eqref{n14}. This is obvious since we are only working with SSM in the background of FNEC. It is clear that to consider SSM in the background of FEC we just need to replace $ a $ with $ a(t) $. Thus, the  SSM metric is embedded in the background of the FLRW cosmology will have the form \eqref{n14} with $ a=a(t) $. We see that using the approach of this section, the metric \eqref{n14} appears very naturally. The calculations in this approach are much simpler than in the previous methods. When $ Q=0 $ (ie $ \overline{Q}=0 $) and $ f(R)=R $ (ie $ M_{ff}=M $) then \eqref{n14} returns to the McVittie metric \cite{McVittie} (can see the McVittie metric in \cite{Arik:2014qua, Nandra:2011ug}). The solution \eqref{n14} coincides with the solution in \cite{Gao:2004cr}. However, to achieve the embedding result, the article \cite{Gao:2004cr} had to require that the metric after embedding must satisfy the equation $G_{01}=0 $ ($ G_{\mu\nu} $ is the Einstein tensor).\\

For a non-static Reissner-Nordström metric, we do the same as above. Perform the coordinate transformations $ x^0=x^0 $, $ \theta=\theta $, $ \varphi=\varphi $, $r=E(q)$ and ignore the second order small quantities. Then the mixing terms of the form $dqdt$ will be ignored. Finally we still have the non-static Reissner-Nordström metric  embedded in the background of the FLRW cosmology has the form \eqref{n14}, where $ \overline{M} $ is replaced by $ \overline{M}(t) $ with $ \overline{ M}(t)=\frac{kc^2M_{ff}(t)}{8\pi} $.\\

The meanings and effects of the metric after embedding can be found in \cite{Kaloper:2010ec, Lake:2011ni, Nandra:2011ug, daSilva:2012nh}. Where, to correct the results in these papers (correct the GR theory), we can replace the $ M $ mass of the GR theory by the $M_{ff}$ effective mass of the $f(R)$ theory.

\section{Uniqueness of solutions of the  $f(R)$ theory and the TOV equation}
In this section, we discuss the uniqueness of solutions to spherically symmetric fields in the  $f(R)$ theory. Specifically, when solving field equations without the presence of a gravitational source $ T^\mu_{~\nu} $ on the right-hand side, the result can have multiple solutions (in some cases, infinitely many solutions). However, more correctly, when the right-hand side of the equations has the presence of $ T^\mu_{~\nu} $, the result will have only one unique solution.
First, for simplicity, we consider the following simple illustrative example: Solve the system of equations
\begin{align}
&B' +A'=\widetilde{T}^0_{~0}, \label{ab1}\\
&\left( A+r\right)\left[  B' + \left(1+\widetilde{T}^0_{~0}\right)A'\right] =\widetilde{T}^0_{~0}\widetilde{T}^1_{~1}. \label{ab2}
\end{align}
Where the functions $ A $ and $ B $ depend on $ r $, the expressions $\widetilde{T}^\mu_{~\nu}$ depend on $ r $ and $T^\alpha_{~\beta}(r)$   such that outside the uncharged gravitational source $T^\alpha_{~\beta}=0$  then $\widetilde{T}^\mu_{~\nu}=0$. It is easy to see that in the absence of gravitational source $ \left( \widetilde{T}^\mu_{~\nu}=0\right) $ the two equations \eqref{ab1} and \eqref{ab2} are 
\begin{align}
&B' +A'=0, \label{abt}\\
&(A+r)(B'+A')=0. \label{abt1}
\end{align}
We see that as long as equation 
\begin{equation}
B'+A'=0 \label{ab3}
\end{equation}
is satisfied, both equations \eqref{abt} and \eqref{abt1} are automatically satisfied. Thus in the absence of a gravitational source $ \left( \widetilde{T}^\mu_{~\nu}=0\right) $, two equations \eqref{ab1} and \eqref{ab2} are not independent. There is only one independent equation, which is  \eqref{ab3}. For each arbitrarily function $A $, solving the equation \eqref{ab3} gives a function $B $. 
That is, the problem will have infinitely many solutions. However, when solving equations in the presence of a gravitational source, the equations \eqref{ab1} and \eqref{ab2} are independent equations. So they will give a unique solution. Indeed, from \eqref{ab1} and \eqref{ab2} we get
\begin{equation}
(A+r)(A'+1)\widetilde{T}^0_{~0}=\widetilde{T}^0_{~0}\widetilde{T}^1_{~1}, \label{ab3a}
\end{equation}
deduce
\begin{equation}
(A+r)(A'+1)=\widetilde{T}^1_{~1}. \label{ab4}
\end{equation}
From \eqref{ab4} we see $ \displaystyle \int (A+r)d(A+r)=\displaystyle \int\widetilde{T}^1_{~1} dr$, it follows that
\begin{equation}
A=-r +\sqrt{2\int^r_0 \widetilde{T}^1_{~1}(r') dr' +(a_0)^2 }. \label{ab5}
\end{equation}
Where $ a_0 $ is a constant. Substituting \eqref{ab5} into \eqref{ab1} we get
\begin{equation}
B=r + \int^r_0\widetilde{T}^0_{~0}(r')dr' -\int^r_0\dfrac{\widetilde{T}^1_{~1}(r')dr'}{\sqrt{\displaystyle\int^{r'}_0 2\widetilde{T}^1_{~1}(x) dx +(a_0)^2 }} +b_0. \label{ab5a}
\end{equation}
Where $ b_0 $ is a constant. The solution \eqref{ab5}-\eqref{ab5a} is general; it can be used both inside and outside the gravitational source. In the case of considering solution outside the gravitational source (in the vacuum), $ r > R_0 $ ($R_0$ is the radius of the gravitational source). Because $ \widetilde{T}^\mu_{~\nu}=0 $ when $ r > R_0 $ then the integrals are only taken between $0$ and $R_0$. Solution \eqref{ab5}-\eqref{ab5a} will become solution in vacuum.
\begin{align}
&A=-r+m_1, \label{ab6}\\
&B=r+m_2. \label{ab7}
\end{align}
Where $m_1$ and $m_2$ are constants, 
\begin{align}
&m_1=\sqrt{2\int^{R_0}_0 \widetilde{T}^1_{~1}(r) dr +(a_0)^2 }, \label{abtm1}\\
&m_2=\int^{R_0}_0\widetilde{T}^0_{~0}(r)dr -\int^{R_0}_0\dfrac{\widetilde{T}^1_{~1}(r)dr}{\sqrt{\displaystyle\int^r_0 2\widetilde{T}^1_{~1}(x) dx +(a_0)^2 }} +b_0. \label{abtm2}
\end{align}
Thus, we see that, in the above example, when solving the system of equations without the presence of the gravitational source, $ \widetilde{T}^\mu_{~\nu}=0 $, the problem will have countless solutions (satisfying equation \eqref{ab3}). However, when $\widetilde{T}^\mu_{~\nu}$ is present in the equations, the result will have only one solution, which is \eqref{ab6}-\eqref{ab7}. The conclusion here is: We usually solve field equations in the absence of a gravitational source, therefore, the problem can have many solutions (see \cite{Xavier:2020ulw, Nguyen:2022bmj, Nguyen:2022blj}). We do not know which of these solutions have physical significance (the others are only mathematically significant). A physically significant solution is one that can be linked to the solution inside the gravitational source. In particular, among the infinite number of solutions satisfying the equation \eqref{ab3} there is only one that has physical significance, that is \eqref{ab6}-\eqref{ab7}. We see \eqref{ab6}-\eqref{ab7} is the solution outside the gravitational source and it can be linked the solution inside the gravitational source to create the general solution \eqref{ab5}-\eqref{ab5a}, which can be used both inside and outside the gravitational source. Thus, to find the physically meaningful solution we must solve the field equations in the presence of the gravitational source $T^\mu_{~\nu}$.\\

In the perturbative approach described in section 4, the general solution that can be used both inside and outside the gravitational source is \eqref{65}--\eqref{68}. Each model $f(R)$ (with known function $f(R)$) will have only one solution. Outside the gravitational source, these solutions become \eqref{38}--\eqref{41}. The uniqueness of the solution has an obvious physical meaning, each gravitational source will only bend space-time in a certain way (there are no two solutions).\\

 In \cite{Xavier:2020ulw}, the authors found the exact solution for a static spherically symmetric field of the model $f(R)=(\alpha_0 + \alpha_1 R)^p $. However, the authors considered only the vacuum condition (without the presence of gravitational source, $T_{\mu\nu} =0$). Thus the authors have obtained infinitely many solutions. Nevertheless, the authors did not know which of these solutions is physically significant. A solution has physical significance if it can be linked to the solution inside the gravitational source to create continuity between solutions. To find the physically meaningful solution we must solve the field equations in the presence of the gravitational source, $T_{\mu\nu}$ is not necessarily zero. If the authors had taken into accountt the presence of $T_{\mu\nu} $, the authors would have obtained a unique solution. A clearer explanation is as follows (notice the equations mentioned in this passage are according to \cite{Xavier:2020ulw}): In \cite{Xavier:2020ulw} we see that, if taking the functions $ A(r) $ and $\delta(r)$ satisfy equation (11) with $ T_2[A(r), \delta(r)] =0 $ then automatically all equations (8a, b , c) are also satisfied. So, from (11) we can choose any function $ \Phi(r) $, then solve the differential equation (11) we will get the function $ A(r) $. Each time we choose a different $ \Phi(r) $ function, we will get a different $ A(r) $ function. Thus, the authors have obtained an infinite number of solutions satisfying the equations (8a, b, c). However, if we now solve equations (8a, b, c) outside the gravitational source (in the vacuum) but in the presence of the gravitational source. That is, there is the presence of $ T_{\mu\nu}(r) $, with $ T_{\mu\nu}(r)=0 $ if $ r>R_0 $, $ T_{\mu\nu} (r)\neq 0 $ if $ r<R_0 $ ($ R_0 $ is the radius of the gravitational source). In this case the right side of equation (9) (so is the left side) will have the components of the tensor $ T_{\mu\nu} $. Hence $T_1[A(r), \delta(r)] \neq 0$ and $T_2[A(r), \delta(r)] \neq 0$. Even expressions of $ T_1 $ and $ T_2 $ contain elements of the tensor $T_{\mu\nu}$. Equations (17) contain components of the tensor $T_{\mu\nu}$ both inside and outside of $T_2$. Then, if we take the functions $A(r) $ and $ \delta(r)$ that only satisfy equation (11) (or (9)), they can not satisfy the equations (8a, b, c) (notice the right side of equations (8a, b, c) now contain the components of the tensor $ T_{\mu\nu} $). Thus in this case, among the equations (8a, b, c) there will be two independent equations of the functions $A(r) $ and $ \delta(r) $ (unlike the case without the presence of $T_{\mu\nu} $, where is only one independent equation, that is equation (11) with $T_2[A(r), \delta(r)] =0$). So in this case we have to solve two independent equations with two unknowns $A(r) $ and $ \delta(r) $. The way to solve equations (8a, b, c) with the right side containing the components of the tensor $ T_{\mu\nu} $ is that we have to integrate the components of tensor $T_{\mu\nu}$. \\
 
In \cite{Numajiri:2021nsc}, the authors give the exact equations describing inside of the gravitational source. In \cite{Numajiri:2021nsc}, if we combine equations (9), (18) and (22), we get the differential equation describing the dependence of the pressure $ P $ on the density of matter $ \rho $ (the TOV equation of the theory $ f(R) $). Combining this TOV equation with equation (9), we get the unique solution of $ g_{00}(r) $. Then using (18), we get the unique solution of $ g_{11}(r) $. That is, inside the gravitational source, the metric tensors $ g_{\mu\nu} $ have unique solution. From this we can conclude that inside of the gravitational source, each of the $f(R)$ theory has only one solution.  Since there is only one solution inside the gravitational source, there is only one solution outside the gravitational source. The solution outside and inside the gravitational source link together to produce a general solution, which is unique and it is a solution that can be described both inside and outside the gravitational source. The uniqueness of the solution has the physical meaning that each known gravitational source only bends space-time in a certain way (only one solution). However, it should be added that the conclusion about the uniqueness of solutions outside the gravitational source (in the vacuum) that we have discussed in this section is not rigorous. A more rigorous proof is left for future work, this is probably a new topic for us to study further.\\

Now let us find the TOV equation of the perturbative $ f(R) $ theory. As mentioned above, in \cite{Numajiri:2021nsc}, the authors have found the exact TOV equation. However, this equation is too complicated and it is not explicit (it does not express the relationship between the pressure $ P $ and the density of matter $ \rho $ directly). This equation also cannot be used in the case of a charged gravitational source. By using the perturbative approach, we can find the perturbative TOV equation, which has a much simpler form. Indeed, the perturbative equation \eqref{9} can be rewritten as 
\begin{align}
R^{\mu}_{~\nu}-\frac{1}{2}\delta^{\mu}_{~\nu}R=-kT^{\mu}_{(eff)\nu}. \label{tov1}
\end{align}
Where
\begin{align}
T^{\mu}_{(eff)\nu}=&T^{\mu}_{~\nu}-\lambda h'(kT)\left(T^{\mu}_{~\nu}
-\frac{1}{2}\delta^{\mu}_{~\nu}T\right)  -\frac{\lambda}{2k}\delta^{\mu}_{~\nu}h(kT) \nonumber\\
&-\frac{\lambda}{k} \delta^{\mu}_{~\nu}\square^E h'(kT)+\frac{\lambda}{k} \nabla^{\mu}\nabla_{\nu}^Eh'(kT) \label{tov2}
\end{align}
is called the effective energy-momentum tensor and by \eqref{me1} and \eqref{fr1}, $ T^{\mu}_{~\nu}= T^{\mu}_{(m)\nu} + T^{\mu}_{(e)\nu} $. We can also rewrite \eqref{66} as
\begin{align}
g_{11}=\frac{-1}{1-\frac{kc^2M_{eff}(r)}{4\pi r}}. \label{tov3}
\end{align}
Where
\begin{align}
M_{eff}(r)=&\left\{\frac{4\pi}{kc^2}\int_0^r\left[ kT^0_{~0}-\frac{\lambda}{2}h(kT^{(m)})
\right.\right.\nonumber \\
&-\lambda k\left( T^0_{~0}-\frac{T^{(m)}}{2}\right) h'(kT^{(m)})\nonumber \\ 
&\left.\left.-\lambda \nabla^i \nabla_i^E h'(kT^{(m)})\right] r'^2dr'\right\}^{-1}, \label{tov4}
\end{align}
with $ k=8\pi G/c^4 $ and $ T^0_{~0}=T^0_{~0}(r', t) $. The perturbative equation \eqref{tov1} differs from the Einstein equation only in that the energy-momentum tensor $ T^\mu_{~\nu} $ is replaced by the effective energy-momentum tensor $ T^\mu_{(eff)\nu} $. From there, we can also easily find the perturbative TOV equation in a static spherically symmetric field,
\begin{align}
\frac{dP}{dr}= \frac{\frac{-G}{r^2}\left(\rho_{eff}+\frac{P_{eff}}{c^2} \right) \left(M_{eff}+\frac{4\pi r^3P_{eff}}{c^2}\right)}{1-\frac{2GM_{eff}}{c^2r}} \label{tov5}
\end{align} 
with
\begin{align}
&T^0_{(eff)0}=\rho_{eff} c^2, \label{tov6}\\
&T^i_{(eff)j}=-P_{eff}\delta^i_j, \label{tov7}\\
&T^0_{~0}=\rho c^2, \label{tov8}\\
&T^i_{~j}=-P\delta^i_j. \label{tov9}
\end{align}
Note that $ T^{\mu}_{(eff)\nu} $ is related to $ T^{\mu}_{~\nu} $ by \eqref{tov2}. Where by \eqref{me1} and \eqref{fr1}, $ T^{\mu}_{~\nu}= T^{\mu}_{(m)\nu} + T^{\mu}_{(e)\nu} $. That is, $ T^{\mu}_{~\nu} $ is the sum of the matter energy-momentum tensor $ T^{\mu}_{(m)\nu} $ and the electromagnetic field energy-momentum tensor $ T^{\mu}_{(e)\nu} $. The \eqref{tov5} equation is the TOV equation of a charged gravitational source in the perturbative f(R) theory. It can be used for any $f(R)$ model. A more detailed study of the application of this equation to the study of stars will be left for future works. However, for the $f(R)=R + \lambda R^2 $ model and for an uncharged gravitational source, we can refer to \cite{Nobleson:2022giu}.

\section{Conclusions}
In the GR theory, the solution outside a spherically symmetric charged gravitational source is \eqref{k1}. Just like the uncharged gravitational source, the metric outside the charged gravitational source is time-independent (stationary state). In other words, when the gravitational source is charged, Birkhoff's theorem does still hold. But inside the gravitational source, the metric tensors can be time-dependent (see \eqref{Aa8} and \eqref{Aa9} in appendix).  $M_f$ is a constant, it is the sum of the masses of the matter, the gravitational field and the electromagnetic field. The value of $M_f$ depends on the radius of the gravitational source and the charge distribution inside the gravitational source. From \eqref{k1}, an investigation of the gravitational acceleration caused by a charged gravitational source shows that the gravitational field is always attractive ($ a_g > 0 $ $\forall r > R_0 $). The event horizon of a black hole was investigated. This shows that the charged gravitational source also has only one  event horizon, which is $r_{sq}$. This radius is larger than the event horizon of the Schwarzschild black hole, $r_{sq} > r_s$.\\

In the perturbative-$f(R)$ theory, the general spherically symmetric solution of a charged gravitational source has the form \eqref{65}--\eqref{68}. The TOV equation of a charged gravitational source is of the form \eqref{tov5}. Outside the gravitational source, these solutions become \eqref{38}--\eqref{41}. Unlike in the GR theory, these solutions  depend on the time $t$ if the radius of the gravitational source $R_0(t)$ depends on time $ t $ (the radius of the gravitational source expands or shrinks). Birkhoff's theorem is broken. This effect makes it possible for a spherically symmetric gravitational source to emit gravitational waves (while the GR equation shows that a spherically symmetric field cannot emit gravitational waves). However, a spherically symmetric field still does not emit electromagnetic waves. To illustrate the solution \eqref{38}--\eqref{41}, we consider the model $ f(R)=R + \lambda R^b $  ($b>0$), then the solutions have the form \eqref{50}--\eqref{53}.  For the model $ f(R)=R -2\Lambda + \gamma R^b $ ($b>0$), the solutions are \eqref{iii1}--\eqref{iii4}.  The solution of the perturbative-$f(R)$ theory corrects  the solution of the GR theory  in terms $\lambda M_1(t)$, $\lambda M_2(t)$ and $\lambda{M_3}(t)$. It means that $ M_f $ is replaced by $ M_{ff} $, with $ M_{ff} $ depends on the time $ t $. From \eqref{50}--\eqref{53bfr} we see that, in comparison  with the uncharged gravitational source, when the gravitational source is charged, there is an additional term $ \dfrac{kQ^2}{32  \pi^2\varepsilon_0r^2} $ in the solution. When $Q=0 $ the term $ \lambda{M_3}(t) $ is zero, but the terms $ \lambda M_1(t) $ and $ \lambda M_2(t) $ are non-zero.  More importantly, the term $ Q $ appears not only in  $ \lambda{M_3}(t) $ but also in  $\lambda M_2(t)$ (see $ \alpha(t) $). Also, the term $Q$  appears in the term $ M_f $. \\[4mm]
A special case with the model $f(R)= R + \frac{\lambda}{R} $, which means $ b=-1 $, we see $ \lambda M_1 =0 $. On the other hand, when the field is static,  we also have $ \lambda M_2 =0 $. Thus, when considering the static field and if the gravitational source is not charged ($ Q=0 $ so $ \lambda {M_3}=0 $), the perturbative solution of the model $f(R)= R + \frac{\lambda}{R} $  will not correct  the solution of the GR equation. The correction occurs only if the gravitational source is charged (because $Q\neq 0 $, so $\lambda{M_3}\neq 0 $. As said, $Q$ also appears in the term $ M_f $.)\\[4mm]
In all cases, the  term $ Q $ appears in the form $ Q^2 $,  which means that the solutions obtained will not depend on the sign of $ Q $. This is obvious because the gravitational equations depend only on the energy-momentum tensor $T^\mu_{~\nu}$. They depend on the energy-momentum tensor of the electromagnetic field, where both  the negative charge and positive charge play the same role.\\

The spherically symmetry metric embedding of a star (or a black hole) in the background of the flat FLRW  cosmological metric  has the form \eqref{n14} (we can also do similar calculations to find the embedding of the \eqref{iii1}-\eqref{iii4} solutions  in the background of the cosmology). This article showed a very simple embedding method and the result appears very naturally.\\

For the static spherically symmetric field, although the perturbative-$f(R)$ theory gives only a small correction to the GR theory (see the event horizon of the Sgr A* black hole in the formula \eqref{rsf}), it is sufficient to correct the physical quantities for greater accuracy with the observations. As an example, it will correct the precession of Mercury's orbit around the Sun or that of the star S2 orbiting the black hole Sgr A* (see the conclusions section of \cite{Ky:2018fer}). The magnitude of the correction term depends on whether we choose a large or small value of $ \lambda $ (as long as the perturbation condition is satisfied). For the non-static spherically symmetric field, the perturbative-$f(R)$ theory will make a significant contribution to the GR theory. For example, a spherically symmetric field can radiate gravitational waves which the GR theory cannot give this effect. Applying the $g_{\mu\nu}$ tensor metrics of the perturbative-$f(R)$ theory, which were found above, we found that, suppose a massive star is collapsing, when the radius of the star approaches the event horizon, gravitational waves burst out very strongly (see \cite{Ky:2024lce}). The perturbative-$f(R)$ theory also leads to some interesting effects in cosmology (see details in \cite{VanKy:2022itq}). Thus, this theory will describe astronomical and cosmological quantities more accurately than Einstein's equation. We hope that this method can be applied to other theories, not just the $f (R)$ theory. This perturbation method will contribute to finding a perfect $f(R)$ model in the future.\\

The uniqueness of the solutions of the $ f(R) $ theory  is discussed in the 6 section. This is probably a new topic for us to study further.

\section*{Acknowledgement} 
This work is supported by the International Physics Centre, Insitute of Physics with grant number  ICP.2022.16.

\appendix
\section{The metric tensors of the GR theory inside a charged gravitational source}

We prove the formula \eqref{36}. 
Using \eqref{65} and \eqref{66}, in the GR limit ($ \lambda h$ is zero) 
we have
\begin{align}
g_{11}(r,t)=\frac{-1}{1-\frac{1}{r}\int_0^r kT^0_{~0}(r', t) r'^2dr'}, \label{Aa1}
\end{align}
\begin{align}
&g_{00}(r,t)=\nonumber\\
&\mbox{exp} \int_\infty^r\left\lbrace r'g_{11}(r',t)\left[ -\frac{1}{r'^2}+kT^{1}_{~1}(r', t)\right] 
-\frac{1}{r'}\right\rbrace dr.'  \label{Aa2}
\end{align}
From \eqref{Aa2}, we obtain
\begin{align}
&g_{00}(r,t)=\nonumber\\
&\exp \int_{R_0(t)}^r\left\lbrace r'g_{11}(r',t)
\left[ -\frac{1}{r'^2}+kT^{1}_{~1}(r', t)\right] -\frac{1}{r'}\right\rbrace dr'\nonumber \\
& \times \exp\int^{R_0(t)}_\infty \left\lbrace r'g_{11}(r',t)
\left[ -\frac{1}{r'^2}+kT^{1}_{~1}(r', t)\right] -\frac{1}{r'}\right\rbrace dr'. \label{Aa2a}
\end{align}
On the other hand, from \eqref{me13} and \eqref{k1}, outside the gravitational source we have $ T^1_{~1}=T^{1(m)}_{~1}+T^{1(e)}_{~1}=T^{1(e)}_{~1}= \frac{1}{2}\varepsilon_0\left(\frac{Q}{4\pi \varepsilon_0 r^2} \right)^2$ and $ g_{11}=-\left( 1-\frac{2GM_f}{c^2r}+\frac{{G}Q^2}{4\pi\varepsilon_0c^4r^2}\right)^{-1} $, hence, 
\begin{align}
&g_{00}(r,t)=\nonumber\\
&\exp \int_{R_0(t)}^r\left\lbrace r'g_{11}(r',t)\left[ -\frac{1}{r'^2}+kT^{1}_{~1}(r', t)\right] 
-\frac{1}{r'}\right\rbrace dr'
\nonumber \\
&\times \left[ 1-\frac{2GM_f}{c^2R_0(t)}+\frac{{G}Q^2}{4\pi\varepsilon_0c^4[R_0(t)]^2}\right]. \label{Aa2c}
\end{align}
From \eqref{Aa1} and  \eqref{Aa2c}, it follows that
\begin{align}
&g_{00}(r,t)=\left[ 1-\frac{2GM_f}{c^2R_0(t)}+\frac{{G}Q^2}{4\pi\varepsilon_0c^4[R_0(t)]^2}\right]
\nonumber \\
&\times \exp\int_{R_0(t)}^r 
\frac{\frac{k}{r'^2}\int_0^{r'} r''^2T^{0}_{~0}(r'',t)dr'' -kr'T^{1}_{~1}(r', t)}
{1-\frac{k}{r'}\int_0^{r'} r''^2T^{0}_{~0}(r'',t)dr''}dr'. \label{Aa3}
\end{align}
Inside the gravitational source $ T^{0(m)}_{~0}\gg T^{0(e)}_{~0} $ so we can take $ T^0_{~0}\simeq T^{0(m)}_{~0} $. On the other hand  $ T^1_{~1}\ll T^0_{~0} $ so \eqref{Aa3} can take  the form
\begin{align}
&g_{00}(r,t)=\left[ 1-\frac{2GM_f}{c^2R_0(t)}+\frac{{G}Q^2}{4\pi\varepsilon_0c^4[R_0(t)]^2}\right]
\nonumber \\
&\times \exp\int_{R_0(t)}^r 
\frac{\frac{k}{r'^2}\int_0^{r'} r''^2T^{0(m)}_{~0}(r'',t)dr''}
{1-\frac{k}{r'}\int_0^{r'} r''^2T^{0(m)}_{~0}(r'',t)dr''}dr'. \label{Aa3a}
\end{align}
We consider  $ T^{0(m)}_{~0} $ depending  on time $ t $ only 
(the density  is considered homogeneous), therefore, 
\begin{align}
&g_{00}(r,t)=\left[ 1-\frac{2GM_f}{c^2R_0(t)}+\frac{{G}Q^2}{4\pi\varepsilon_0c^4[R_0(t)]^2}\right]\nonumber\\
&\times\exp\int_{R_0(t)}^r 
\frac{\frac{k}{3}r'T^{0(m)}_{~0}(t) }{1-\frac{k}{3} r'^2T^{0(m)}_{~0}(t)}dr'. \label{Aa4}
\end{align}
From \eqref{Aa4} we obtain 
\begin{align}
g_{00}(r,t)=&\left[ 1-\frac{2GM_f}{c^2R_0(t)}+\frac{{G}Q^2}{4\pi\varepsilon_0c^4[R_0(t)]^2}\right]\times\nonumber\\
&\sqrt{\frac{1-
\frac{k}{3}[R_0(t)]^2T^{0(m)}_{~0}(t)}{1-\frac{k}{3}r^2T^{0(m)}_{~0}(t)}}. \label{Aa6}
\end{align}
Substituting
\begin{equation}
T^0_{~0}\simeq T^{0(m)}_{~0}(t) = \frac{Mc^2}{\frac{4}{3}\pi [R_0(t)]^3}
\label{tt}
\end{equation}
for \eqref{Aa6} and \eqref{Aa1},  we have the tensor metrics inside the charged gravitational source of the GR theory
\begin{align}
g_{00}(r,t)=&\left[ 1-\frac{2GM_f}{c^2R_0(t)}+\frac{{G}Q^2}{4\pi\varepsilon_0c^4[R_0(t)]^2}\right]\nonumber\\
&\times\sqrt{\frac{1-\frac{kMc^2}{4\pi R_0(t)}}{1-\frac{kMc^2r^2}{4\pi [R_0(t)]^3}}}, \label{Aa8}
\end{align}
\begin{align}
g_{11}(r,t)=\frac{-1}{1-\frac{kMc^2r^2}{4\pi [R_0(t)]^3}}. \label{Aa9}
\end{align}
From \eqref{Aa8} and \eqref{Aa9}, we see that, if the radius of the gravitational source $R_0(t)$ changes on time, the tensor metrics depend on time $t$.\\ 

Substituting  \eqref{Aa8} and  \eqref{Aa9} into  \eqref{35}, we get 
\begin{align}
&\nabla^i \nabla_i^E h'(kT^{0(m)}_{~0})\cong\nonumber\\
 &\frac{k}{8\pi }
\frac{\frac{\partial h'(kT^{0(m)}_{~0})}{\partial t} }{\left[ 1-\frac{2GM_f}{c^2R_0(t)}+\frac{{G}Q^2}{4\pi\varepsilon_0c^4[R_0(t)]^2}\right] \sqrt{1-\frac{kMc^2}{4\pi R_0(t)}}} \nonumber \\
&\times \frac{r^2}{\sqrt{1-\frac{kMc^2r^2}{4\pi [R_0(t)]^3}}}
\frac{\partial}{\partial t}\ \frac{M}{[R_0(t)]^3} \nonumber\\
=& \frac{k}{8\pi}\frac{\frac{\partial (kT^{0(m)}_{~0})}{\partial t}h''(kT^{0(m)}_{~0})}{\left[ 1-\frac{2GM_f}{c^2R_0(t)}+\frac{{G}Q^2}{4\pi\varepsilon_0c^4[R_0(t)]^2}\right] \sqrt{1-\frac{kMc^2}{4\pi R_0(t)}}}
\nonumber \\
&\times\frac{r^2}{\sqrt{1-\frac{kMc^2r^2}{4\pi [R_0(t)]^3}}}
\frac{\partial}{\partial t} \frac{M}{[R_0(t)]^3}. \label{Aa11}
\end{align}
Then, from  \eqref{tt} and  \eqref{Aa11}, we obtain
\begin{align}
&\nabla^i \nabla_i^E h'(kT^{0(m)}_{~0})\cong\nonumber\\
& \frac{3k^2c^2}{32\pi^2}
\frac{h''(kT^{0(m)}_{~0})}{\left[ 1-\frac{2GM_f}{c^2R_0(t)}+\frac{{G}Q^2}{4\pi\varepsilon_0c^4[R_0(t)]^2}\right] \sqrt{1-\frac{kMc^2}{4\pi R_o(t)}}}\nonumber \\
&\times\frac{r^2}{\sqrt{1-\frac{kMc^2r^2}{4\pi [R_0(t)]^3}}}
\left[ \frac{\partial}{\partial t}\frac{M}{[R_0(t)]^3}\right]^2. \label{Aa12}
\end{align}
It follows that 
\begin{align}
&\int_0^{R_0(t)}\nabla^i \nabla_i^E h'(kT^{0(m)}_{~0})r^2dr\cong \frac{3k^2c^2}{32\pi^2}
\left[ \frac{\partial}{\partial t}\frac{M}{[R_0(t)]^3}\right]^2 \nonumber \\
&\times \frac{h''(kT^{0(m)}_{~0})}{\left[ 1-\frac{2GM_f}{c^2R_0(t)}+\frac{{G}Q^2}{4\pi\varepsilon_0c^4[R_0(t)]^2}\right] \sqrt{1-\frac{kMc^2}{4\pi R_0(t)}}}\nonumber\\
&\times\int_0^{R_0(t)}\frac{r^4}{\sqrt{1-\frac{kMc^2r^2}{4\pi [R_0(t)]^3}}}dr. \label{Aa13}
\end{align}
Set
\begin{align}
\xi^2 (t)=\frac{kMc^2}{4\pi [R_0(t)]^3} \label{Aa14}
\end{align}
we rewrite \eqref{Aa13} as 
\begin{align}
&\int_0^{R_0(t)}\nabla^i \nabla_i^E h'(kT^{0(m)}_{~0})r^2dr\cong \frac{3k^2c^2}{32\pi^2}
\left[ \frac{\partial}{\partial t}\frac{M}{[R_0(t)]^3}\right]^2 \nonumber \\ 
&\times \frac{h''(kT^{0(m)}_{~0})}{\left[ 1-\frac{2GM_f}{c^2R_0(t)}+\frac{{G}Q^2}{4\pi\varepsilon_0c^4[R_0(t)]^2}\right] \sqrt{1-\frac{kMc^2}{4\pi R_0(t)}}}\nonumber\\
&\times\int_0^{R_0(t)}\frac{r^4}{\sqrt{1-\xi^2 (t)r^2}}dr. \label{Aa15}
\end{align}
On the other hand, we can easily see that
\begin{align}
&\frac{r^4}{\sqrt{1-\xi^2r^2}}=\nonumber\\
&\frac{1}{8\xi^5}\frac{\partial}{\partial r}
\left[3\mbox{arcsin}(\xi r)-\xi r(3+2\xi^2 r^2)\sqrt{1-\xi^2r^2} \right]. \label{Aa16}
\end{align}
Thus, 
\begin{align}
&\int_0^{R_0(t)}\nabla^i \nabla_i^E h'(kT^{0(m)}_{~0})r^2dr\nonumber\\
&\cong h''(kT^{0(m)}_{~0})\left[ \frac{\partial}{\partial t}\frac{M}{[R_0(t)]^3}\right]^2 \alpha (t), \label{Aa17}
\end{align}
where
\begin{align}
&\alpha (t)=\frac{3k^2c^2R_0(t)}{256\pi^2\vartheta(t)[\xi (t)]^4}\left\lbrace \frac{3}{\xi(t)R_0(t)}
\arcsin[\xi (t) R_0(t)]\right.\nonumber \\
 &\left. -\left( 3+2[\xi(t)R_0(t)]^2\right) \sqrt{1-[\xi(t)R_0(t)]^2}\right\rbrace, \label{Aa18}
\end{align}
and
\begin{align}
\vartheta(t)=&\left( 1-\frac{2GM_f}{c^2R_0(t)}+\frac{{G}Q^2}{4\pi\varepsilon_0c^4[R_0(t)]^2}\right) \sqrt{1-\frac{kMc^2}{4\pi R_0(t)}}\nonumber\\
=&\left( 1-\frac{2GM_f}{c^2R_0(t)}+\frac{{G}Q^2}{4\pi\varepsilon_0c^4[R_0(t)]^2}\right) \sqrt{1-\frac{2GM}{c^2R_0(t)}}. \label{Aa19}
\end{align}

\end{document}